\documentclass[useAMS,usenatbib]{mn2e}
\usepackage{epsfig}
\usepackage{natbib}
\usepackage{lscape}
\usepackage[english]{babel}

\title[Long-term variability of T Tauri stars]{Long-term variability of T Tauri stars using WASP}

\author[Rigon et al.]{Laura Rigon$^{1}$, Alexander Scholz$^{1}$\thanks{corresponding author: as110@st-andrews.ac.uk}, David Anderson$^{2}$, Richard West$^{3}$\\
$^{1}$SUPA, School of Physics \& Astronomy, University of St Andrews, North Haugh, St Andrews, KY16 9SS, United Kingdom\\
$^{2}$Faculty of Natural Sciences, Keele University, ST5 5BG, United Kingdom\\
$^{3}$Department of Physics \& Astronomy, University of Leicester, University Road, Leicester, LE1 7RH, United Kingdom\\
}

\begin{document}

\date{Accepted  Received }


\maketitle

\label{firstpage}

\begin{abstract}
We present a reference study of the long-term optical variability of young stars using data from the WASP project. 
Our primary sample is a group of well-studied classical T Tauri stars (CTTS), mostly in Taurus-Auriga. WASP lightcurves 
cover timescales up to 7 years and typically contain 10000-30000 datapoints. We quantify the variability as function
of timescale using the time-dependent standard deviation 'pooled sigma'. We find that the overwhelming majority of CTTS
has low-level variability with $\sigma < 0.3$\,mag dominated by timescales of a few weeks, consistent with
rotational modulation. Thus, for most young stars monitoring over a month is sufficient to constrain the
total amount of variability over timescales up to a decade. The fraction of stars with strong optical variability 
($\sigma > 0.3$\,mag) is 21\% in our sample and 21\% in an unbiased control sample. An even smaller fraction 
(13\% in our sample, 6\% in the control) show evidence for an increase in variability amplitude as a function 
of timescale from weeks to months or years. The presence of long-term variability correlates with the spectral slope 
at 3-5\,$\,\mu m$, which is an indicator of inner disk geometry, and with the U-B band slope, which is an accretion diagnostics. 
This shows that the long-term variations in CTTS are predominantly driven by processes in the inner disk and in the 
accretion zone. Four of the stars with long-term variations show periods of 20-60\,d, significantly longer than 
the rotation periods and stable over months to years. One possible explanation are cyclic changes in the interaction 
between the disk and the stellar magnetic field.
\end{abstract}


\section{Introduction}

Young stars are variable. Although this has been known for more than half a century, we have only started to gain a comprehensive understanding of the characteristics and the origins of these variations. 
Until the late 1990s, most photometric studies of the variability in young stellar objects (YSOs) were limited to the optical part of the spectrum and to timescales ranging from hours to weeks, focusing on studies of rotation periods due to surface spots and accretion-induced changes \citep[e.g.][]{bouvier93,fernandez96,stassun99,lamm}, see also review by \citet{herbst07}.

With the advent of large-scale infrared surveys via satellites like Spitzer and WISE as well as large-scale variability surveys (e.g., PTF, COROT), our knowledge of YSO variability has improved rapidly over the past decade. The characterisation has been extended to the near- \citep[e.g.][]{carpenter,eiroa02,alves08,rice12}, mid- \citep[e.g.][]{rebull14,morales11,flaherty12}, and even far-infrared \citep{billot12}. Also, the number of sources known to be undergoing phases of exceptional variability, like FU Ori-type accretion bursts or disk-induced eclipses, has increased substantially in the last few years \citep[e.g.][]{miller11,caratti11,rodriguez12,plavchan13}. With the simultaneous monitoring campaign of NGC2264 using optical (COROT) and infrared (Spitzer) space telescopes, the variability of YSOs has been studied with unprecedented photometric precision and cadence, and time coverage from minutes to several months, leading to a new detailed morphological taxonomy \citep{cody14}, newly discovered phenomena \citep{stauffer14}, as well as first physical models for the origin of the variations \citep[e.g.][]{kurosawa,kesseli}. On these relatively short timescales, YSO variability carries signatures of strictly periodic or quasi-periodic modulations through cool and/or hot spots and inner disk features, as well as stochastic variations, either induced by the inner disk or through bursts and flares. In many cases, a mixture of these processes seems to be present, leading to bewilderingly complex lightcurves. 

The behaviour on timescales exceeding $\sim 1$\,year has not been explored yet in the same detailed manner. Notable exceptions include the ROTOR program by \citet{grankin}, and work by \citet{gahm93,percy,ibryamov}. In these publications, the sampling is limited to a few datapoints per night and in total up to $\sim 1000$ datapoints. Our goal in this paper is to extend these long-term studies by analysing several years of high-cadence monitoring for several dozens of T Tauri stars, with data from the exoplanet hunting project Super-WASP. Our time series cover usually 7 years and contain typically several thousand datapoints in white light. Our particular emphasis is to search for links between the presence of long-term variability and characteristics of stars and disks, to constrain the origin of the brightness changes.

\section{Sample and data}
\label{sec:dataprep}

\subsection{Sample}

The aim of this paper is to characterise the long-term behaviour of CTTS in photometric lightcurves and to study possible origins for the variability. Therefore, we focused on a group of 39 well-studied T Tauri stars, listed in Table \ref{t1}. For all these objects, a wealth of information is available about the stellar properties as well as about the disk, which allows us to probe connections between variability and other parameters. These stars are part of the sample of the DIANA project, which aims at analysing protoplanetary disks at all wavelengths, from UV to mm, both from a theoretical and an observational point of view \citep{woitke}. 

The objects in our sample are located in various star forming regions; most of them in Taurus-Auriga (26/39), 7 in $\rho$\,Ophiuchus and Upper Scorpius, 3 in Lupus, and 3 in the TW Hydrae association. The approximate ages of the stars range from 1 to 10\,Myr. Apart from one F- and two G-type stars, the sample is comprised of K- and M-dwarfs and hence dominated by young stars with masses at or below the mass of the Sun. The latest spectral type is M4, corresponding to approximately 0.3-0.4$\,M_{\odot}$. For all objects in our sample there is evidence for the presence of the disk; the overwhelming majority are 'class II'.

Because we focus on well-known T Tauri stars, our sample contains many objects that have been studied over decades and have originally been discovered based on their variability, e.g., RW Aur, T Tau, BP Tau \citep{joy45}. Therefore, the sample may be biased towards highly variable young. To evaluate this possible bias, we select a control sample from the disk survey by \citet{luhman}. They cover 352 objects in Taurus, a large fraction of the total known young population in this region, selected without any bias with respect to variability. Of these objects, 183 are 'class II', i.e. are considered to have a disk. From these 183, 81 have a lightcurve in the WASP database. The spectral type distribution in this unbiased control sample is similar to the primary sample, most stars have ages around 1-2\,Myr. 

\subsection{WASP data}
\label{waspdata}

For all stars in our sample, the database of the WASP project contains a large number of photometric datapoints. The main purpose of WASP is the detection of transits due to exoplanets; to date more than hundred exoplanets have been discovered based on WASP transits \citep[e.g.][]{cameron1,hellier,anderson}. Apart from the exoplanet science, the WASP lightcurves are a useful resource to study other types of variable stars; here we exploit the database to examine the variability of young stars.

The WASP observations are carried out with two robotic telescopes, one on La Palma, the other in South Africa. Each telescope consistes of eight Canon lenses and is equipped with a 2Kx2K CCD, with a field of view of $7.8\times 7.8$\,degrees. This results in a pixel scale of 13.7" pixel$^{-1}$. The observations for the WASP project started in 2004. The transmission of the WASP system is essentially a white filter with cutoffs at 400 and 750\,nm. Typically, the telescope observes 6-9 fields per night, at similar declination and spaced in right ascension. This results in a high cadence over the course of a night. The long-term lightcurves have significant gaps, due to target fields being behind the Sun, technical down time, or observations of other fields. The data is processed with a custom-built pipeline. For more details on the WASP project, see \citet{pollacco}.

For this study, we downloaded the pipeline-produced lightcurves for our target stars from the WASP database. The format of the pipeline output is described in \citet{swasp}; for our purposes we use the flux in column TAMFLUX2, its associated error TAMFLUX2\_ERR, and the mid-time of exposure TMID. TAMFLUX2 is the flux corrected for systematics and is based on the SysRem algorithm (see \citet{cameron2} and \citet{tamuz}). The flux errors correspond to the white noise and do not include additional systematics. We converted the epochs to days after the zero epoch 2004-01-01T00:00:00 and the fluxes to magnitudes, following the description in \citet{swasp}. The flux errors were propagated to magnitude errors:

\begin{equation}
\mathrm{err}=2.5 \cdot \log_{10}{e} \cdot \mathrm{\frac{TAMFLUX2\_ERR}{TAMFLUX2}}
\end{equation}

For our target stars, the lightcurves have at least 1000, and typically 10000 to 30000 datapoints. They cover about 7 years. The median number of datapoints per night varies from 24 to 82 in our sample. The average WASP magnitude and the standard deviation of the lightcurves are listed in Table \ref{ctts}. The lightcurve of the control sample have comparable properties.

\begin{figure}
\includegraphics[scale=0.5]{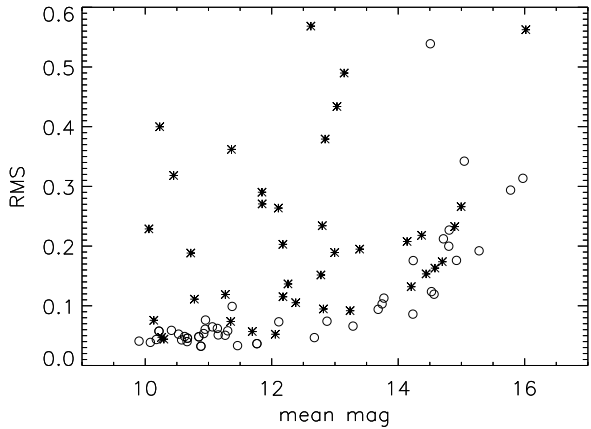}
\caption{Mean WASP magnitude vs. standard deviation for our primary sample of T Tauri stars (asterisks) and for a sample of
random field stars selected from fields around our targets (open circles).}
\label{mag-err}
\end{figure}                          

WASP lightcurves are taken under a range of conditions; the quality of the data is therefore fairly inhomogenuous. To homogenise and clean the lightcurves, we applied a series of filters. Outliers were removed from the sample, using a simple $\sigma$-rejection criterion. Data points were considered outliers in case they are outside $\overline{mag}\pm 6\sigma$. For purely Gaussian noise, we would expect 0 outliers with this criterion, but we find 0 to 147 in our sample. In addition, we retained only datapoints with errors lower or equal to 1\,mag. A visual comparison of the lightcurves of targets revealed that the noise was significantly enhanced in 18 nights for stars in the northern hemisphere. These 'bad' nights are irregularly spread over all seasons and are presumably due to suboptimal observing conditions. We do not use data from these nights for stars in the north for all further analysis.

For the two stars LkHa326 and LkHa327 we removed the data taken from September 2007 to February 2008. During that timespan, these two objects, 52.18' distant from each other and therefore measured on the same images at the same epochs, show some similarities in their lightcurves. In particular there are great variations in magnitude between September and December 2007, and both stars appear to be a magnitude fainter after that. Other stars in the area of the sky between these two sources show the same characteristics. Although we cannot reliably ascertain the nature of these changes, they seem to be due to instrumental or atmospheric effects and hence the data not useful for our purposes. 

For stars with WASP magnitudes ranging from 10 to 14, the photometric errors taken from the WASP database are in the range of 1-5\%, for our primary and for the comparison sample. As noted above, these values do not include possible systematics which may be substantial over the long timescales considered here. To estimate realistic errors, we selected a random set of reference stars located close to stars from our primary sample. Most of these stars should be non-variable, i.e. the standard deviation of their lightcurves provides an estimate of the full photometric error. The standard deviations for reference stars of the same magnitude, but from different fields, are very similar, i.e. we can use the total sample of reference stars to derive errors for all our targets. 

In Fig. \ref{mag-err} we show the RMS of the lightcurves in our primary sample (asterisks) vs. the RMS of the reference star lightcurves (empty circles). The floor of the datapoints for the reference stars should define the photometric error as a function of magnitude. From this plot, it is clear that the overwhelming majority of our targets have an RMS significantly enhanced compared to the photometric error, i.e. they show evidence for variability. We confirm this by comparing the standard deviation in target lightcurves with the corresponding photometric error using the F-test. For 26/39 objects, this yields $F>3$, corresponding to a likelihood $>99\%$ that the noise in their lightcurves is larger than the error. All these objects are significantly variable compared with the photometric errors. For only a handful of objects the standard deviation in the lightcurve is consistent with the errors ($F\sim 1$). 


\begin{table*}
\caption{Target sample with coordinates (J2000), spectral type, disk inclination $i$, multiplicity $m$, average WASP magnitude and
standard deviation, reference for spectral type, inclination, and binarity. The second column indicates if the star belongs
to group A or B (see Sect. \ref{subsec:PooledSigma}). For the inclination, we add a code to indicate the method used in the
literature: R ($v\sin{i}$ and rotation period), S (submm/mm), I (infrared), O (optical observations).
\label{t1}}                                   
\center
\begin{tabular} {@{}lccccccccccc@{}}
\hline
Name        &   Group  & $\alpha$ & $\delta$       & Sp.Type  & $i$ (method) & $m$ &$\overline{mag}\pm \sigma$ & Ref\\
            &          & h m s    & $^\circ$ ' ''  &          & deg  &     & mag                       &(SpT,incl,binarity)\\      
\hline
AATau            &A & 04 34 55.42  & +24 28 53.20   & M0.6  & 75 (O) &  1   &   12.85 $\pm$ 0.38 &   a,1,15    \\     
BPTau            &- & 04 19 15.84  & +29 06 26.90   & M0.5  & 39 (T) &  1   &   12.26 $\pm$ 0.14 &   b,2,15    \\
CITau            &- & 04 33 52.00  & +22 50 30.20   & K5.5  & 46 (S) &  1   &   12.78 $\pm$ 0.15 &   a,3,15    \\
CQTau            &A & 05 35 58.47  & +24 44 54.10   & F3    & 66 (O) &  1   &   10.23 $\pm$ 0.40 &   c,4    \\
CWTau            &B & 04 14 17.00  & +28 10 57.83   & K3    &        &  1   &   12.62 $\pm$ 0.57 &    d,-,15	\\  
DFTau            &- & 04 27 02.80  & +25 42 22.31   & M2.7  &        &  2   &   11.85 $\pm$ 0.27 &    a,-,18    \\
DNTau            &- & 04 35 27.37  & +24 14 58.90   & M0.3  & 77 (S) &  1   &   12.38 $\pm$ 0.11 &  a,5,15     \\
DoAr24E          &- & 16 26 23.35  & $-$24 20 59.78 & K0    &        &  2   &   14.44 $\pm$ 0.15 &    e,-,17	\\
DOTau            &A & 04 38 28.58  & +26 10 49.40   & M0    & 42 (S) &  1   &   13.03 $\pm$ 0.43 &   d,5,15    \\
DRTau            &- & 04 47 06.21  & +16 58 42.80   & K6    & 67 (S) &  1   &   11.85 $\pm$ 0.29 &   a,5    \\	
FSTau            &- & 04 22 02.18  & +26 57 30.49   & M2.4  &        &  2   &   14.70 $\pm$ 0.17 &   a,-,16    \\  
FTTau            &- & 04 23 39.19  & +24 56 14.11   & M2.8  & 60 (S) &  1   &   14.37 $\pm$ 0.22 &   a,3  \\ 
GGTau            &- & 04 32 30.35  & +17 31 40.60   & K7.5  & 37 (S) &  2   &   12.18 $\pm$ 0.12 &  a,6,16   \\ 
GOTau            &- & 04 43 03.09  & +25 20 18.80   & M2.3  & 66 (S) &  1   &   14.89 $\pm$ 0.23 &  a,3,15   \\  
GQLup            &- & 15 49 12.14  & $-$35 39 03.95 & K5    &        &  1   &   11.27 $\pm$ 0.12 &  a	 \\  
Haro1-16         &- & 16 31 33.46  & $-$24 27 37.30 & K3    &        &  1   &   12.82 $\pm$ 0.09 &   e	 \\
Haro6-13         &B & 04 32 15.41  & +24 28 59.70   & M0    &        &  1   &   16.02 $\pm$ 0.56 &   f    \\
Hen3-600A        &- & 11 10 27.88  & $-$37 31 52.00 & M4.1  &        &  2   &   12.06 $\pm$ 0.05 &   a,-,19    \\
HKTauB           &- & 04 31 50.57  & +24 24 18.10   & M1.5  & 85 (I) &  2   &   15.00 $\pm$ 0.27 &  a,7,16   \\    
HLTau            &- & 04 31 38.43  & +18 13 57.60   & K3    & 53 (S) &  3   &   12.78 $\pm$ 0.23 &   a,5,20  \\   
HTLup            &- & 15 45 12.87  & $-$34 17 30.59 & K2    &        &  3   &   10.26 $\pm$ 0.05 &   a,-,16    \\ 
HVTauC           &- & 04 38 32.00  & +26 11 00.00   & M1    & 84 (I) &  1   &   14.21 $\pm$ 0.13 &   q,8  \\   
IMLup            &- & 15 56 09.22  & $-$37 56 05.80 & K6    & 54 (S) &  1   &   11.35 $\pm$ 0.07 &   a,9  \\
IQTau            &B & 04 29 51.56  & +26 06 44.90   & M1.1  & 71 (S) &  1   &   13.15 $\pm$ 0.49 &   a,5,15  \\
IRAS04189+2650   &- & 04 22 00.70  & +26 57 32.5    & K5    &        &  1   &   14.58 $\pm$ 0.16 &   f    \\ 
LkHa326          &- & 03 30 44.06  & +30 32 46.95   & M0    &        &  1   &   12.98 $\pm$ 0.53 &   g    \\
LkHa327          &- & 03 33 30.42  & +31 10 50.40   & K2    &        &  1   &   14.11 $\pm$ 0.35 &   h    \\
RULup            &A & 15 56 42.31  & $-$37 49 15.50 & M0    & 24 (R) &  1   &   11.36 $\pm$ 0.36 &   i,10 \\
RWAur            &A & 05 07 49.54  & +30 24 05.07   & K6.5  & 77 (I) &  3   &   10.45 $\pm$ 0.32 &    a,11,16\\
RYTau            &- & 04 21 57.40  & +28 26 35.54   & G0    & 66 (S) &  1   &   10.06 $\pm$ 0.23 &   a,12 \\
TTauN            &- & 04 21 59.42  & +19 32 06.48   & K0    & 20 (I) &  1   &   10.14 $\pm$ 0.08 &   a,22 \\
TWA07            &- & 10 42 30.11  & $-$33 40 16.20 & M3.2  &        &  1   &   10.72 $\pm$ 0.19 &   a-,21    \\
TWHya            &- & 11 01 51.92  & $-$34 42 17.00 & M0.5  & 6 (S)  &  1   &   10.78 $\pm$ 0.11 &    a,13\\
USCoJ1604-2130   &- & 16 04 21.66  & $-$21 30 28.40 & K2    &        &  -   &   12.11 $\pm$ 0.26 &   l    \\  
UZTauE           &- & 04 32 43.04  & +25 52 31.10   & M1    & 54 (S) &  2   &   12.18 $\pm$ 0.20 &   d,14,16 \\
V1121Oph         &- & 16 49 15.30  & $-$14 22 08.63 & K4    &        &  1   &   11.69 $\pm$ 0.06 &   m    \\
V1149Sco         &- & 15 58 36.92  & $-$22 57 15.30 & G7    &        &  1   &   10.29 $\pm$ 0.04 &   n    \\ 
V853Oph          &- & 16 28 45.28  & $-$24 28 19.00 & M1.5  &        &  2   &   13.39 $\pm$ 0.19 &   o,-,17    \\ 
WaOph6           &- & 16 48 45.63  & $-$14 16 35.96 & K7    & 41 (S) &  1   &   13.24 $\pm$ 0.09 &   p,3  \\
\hline                                                                                     
\end{tabular}       
                                                                       
\small{(1) \cite{bouvier}, (2) \cite{muzerolle2003}, (3) \cite{andrews2007b}, (4) \cite{natta}, (5) \cite{kitamura}, (6) \cite{pietu}, (7) \cite{mccabe}, (8) \cite{monin}, (9) \cite{panic}, (10) \cite{stempels}, (11) \cite{eisner}, (12) \cite{isella}, (13) \cite{qi}, (14) \cite{simon}, (15) \cite{white-ghez}, (16) \cite{woitas}, (17) \cite{barsony}, (18) \cite{tamazian}, (19) \cite{correia}, (20) \cite{welch}, (21) \cite{muzerolle2001}, (22) \cite{ratzka}, (a) \cite{herczeg2014}, (b) \cite{andrews2007b}, (c) \cite{hernandez}, (d) \cite{kenyon}, (e) \cite{bouvierSpt}, (f) \cite{luhman}, (g) \cite{casali}, (h) \cite{fernandez}, (i) \cite{hughes}, (l) \cite{kohler}, (m) \cite{torres}, (n) \cite{houk}, (o) \cite{cohen}, (p) \cite{grankin}}\\
\label{ctts}                                                                               
\end{table*}

\section{Lightcurve analysis}

The WASP lightcurves of our primary sample show a great deal of variety and complexity. Visual examination indicates that regular variations which appear to be periodic are combined with irregular variations and slow trends, as expected for CTTS. The comparison of the lightcurve standard deviation with the photometric errors indicates significant variability for the clear majority stars (see Sect. \ref{waspdata}). In the following, we quantify the variability as a function of timescale.

\subsection{Pooled Sigma}
\label{subsec:PooledSigma}

To broadly characterise the variability, we use a metric called 'pooled sigma' in the following. To derive the pooled sigma, a lightcurve is divided in time windows of a defined length $\Delta t$ ('bins'). For each bin, the standard deviation $\sigma_i$ is calculated. The pooled sigma is then the average of these values. This yields information about the variations in the lightcurve as a function of $\Delta t$. We combine this with appropriate weighting and error analysis. A similar method with variance instead of standard deviation was used by \citet{dobson} and \citet{scholz}. For a variable star, the pooled sigma is expected to rise with $\Delta t$ and plateau around typical timescales of the variability (for example, the rotation period). For the complex lightcurves of CTTS and the irregular sampling of WASP, this turns out to be a robust way to characterise the involved timescales, without assuming anything about the specific type of variability. Pooled sigma is a quantity that is related to and yields similar results as various other metrics used in the literature, such as the $\Delta m$-$\Delta t$ plots discussed by \citet{findeisen}, the self-correlation diagram by \citet{percy2}, and can also be compared with the structure function often used in analysis of AGN lightcurves \citep{emmanoulopoulos}.

The pooled sigma in this paper was calculated as follows. 
\begin{equation}\label{pooledsigma_eq}
\sigma_{pool}=\sqrt{\frac{\sum_{i=1}^{n_{bins}}{(N_i-1)\sigma_i^2}}{\sum_{i=1}^{n_{bins}}{N_i-1}}}
\end{equation}
where, for each $i$-bin, $N_i$ is the number of data points and $\sigma_i$ is the standard deviation from the mean, 
\begin{equation}\label{dev}
\sigma_i=\sqrt{\sum_{i=1}^{N}\frac{(x_i-\bar{x})^2}{N-1}}
\end{equation}
 and $\bar{x}$ is the optimal average, weighted with the photometric errors:
\begin{equation}\label{mean}
\bar{x}=\frac{\sum_{i=0}^{N}\frac{x_i}{\sigma_{err_i}^2}}{\sum_{i=0}^{N}\frac{1}{\sigma_{err_i}^2}}
\end{equation}

\begin{figure*}
\includegraphics[scale=0.45]{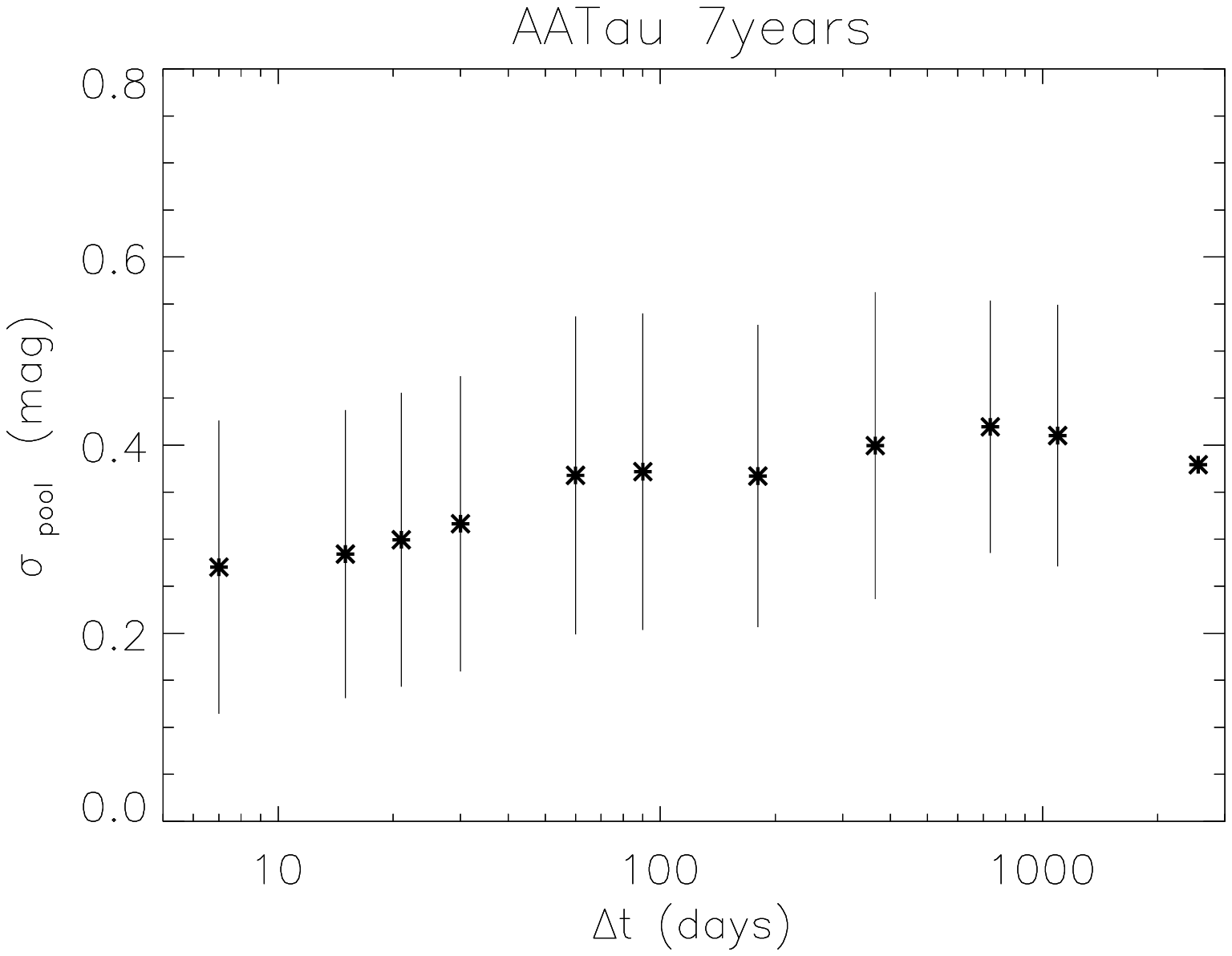}
\includegraphics[scale=0.45]{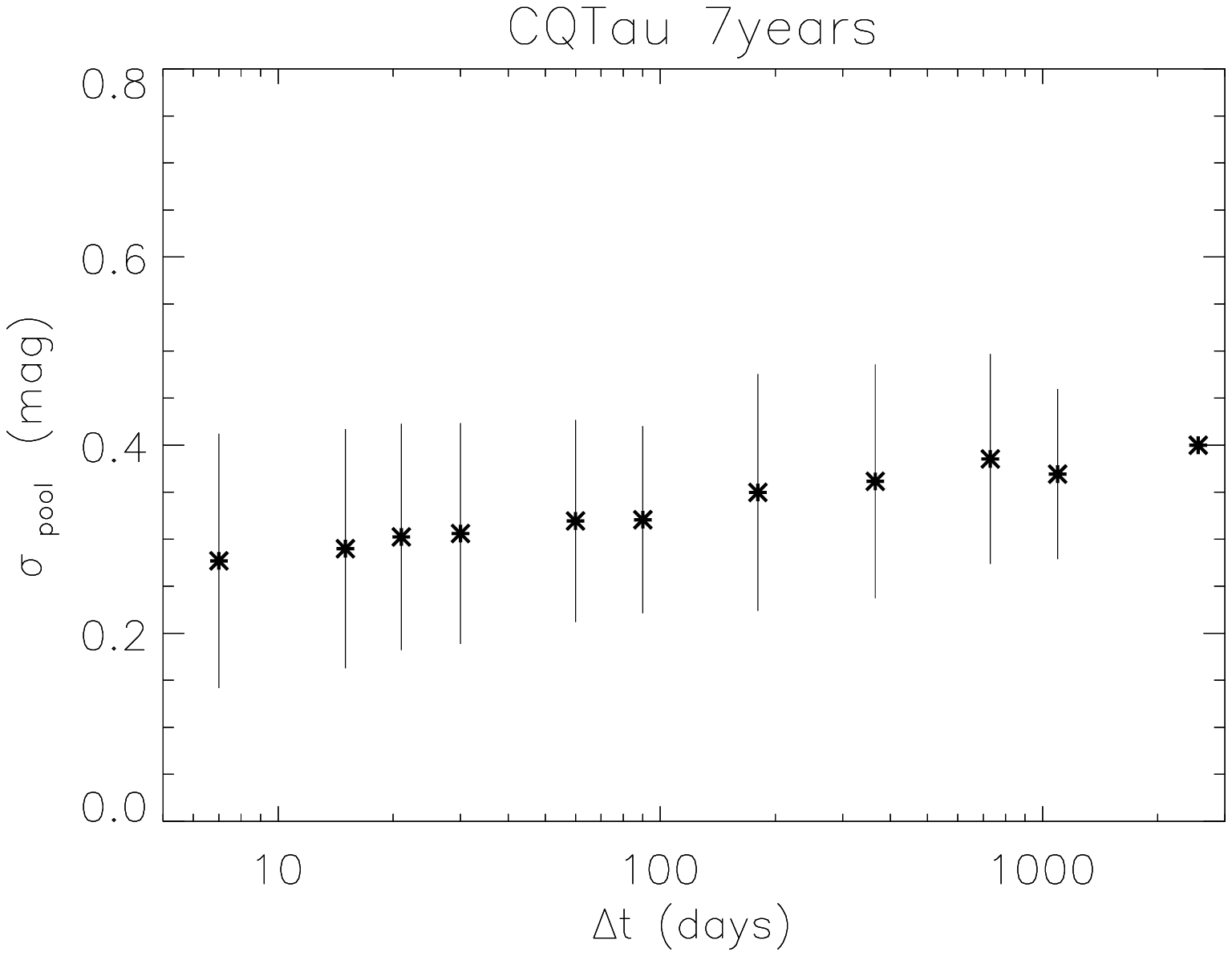}
\includegraphics[scale=0.45]{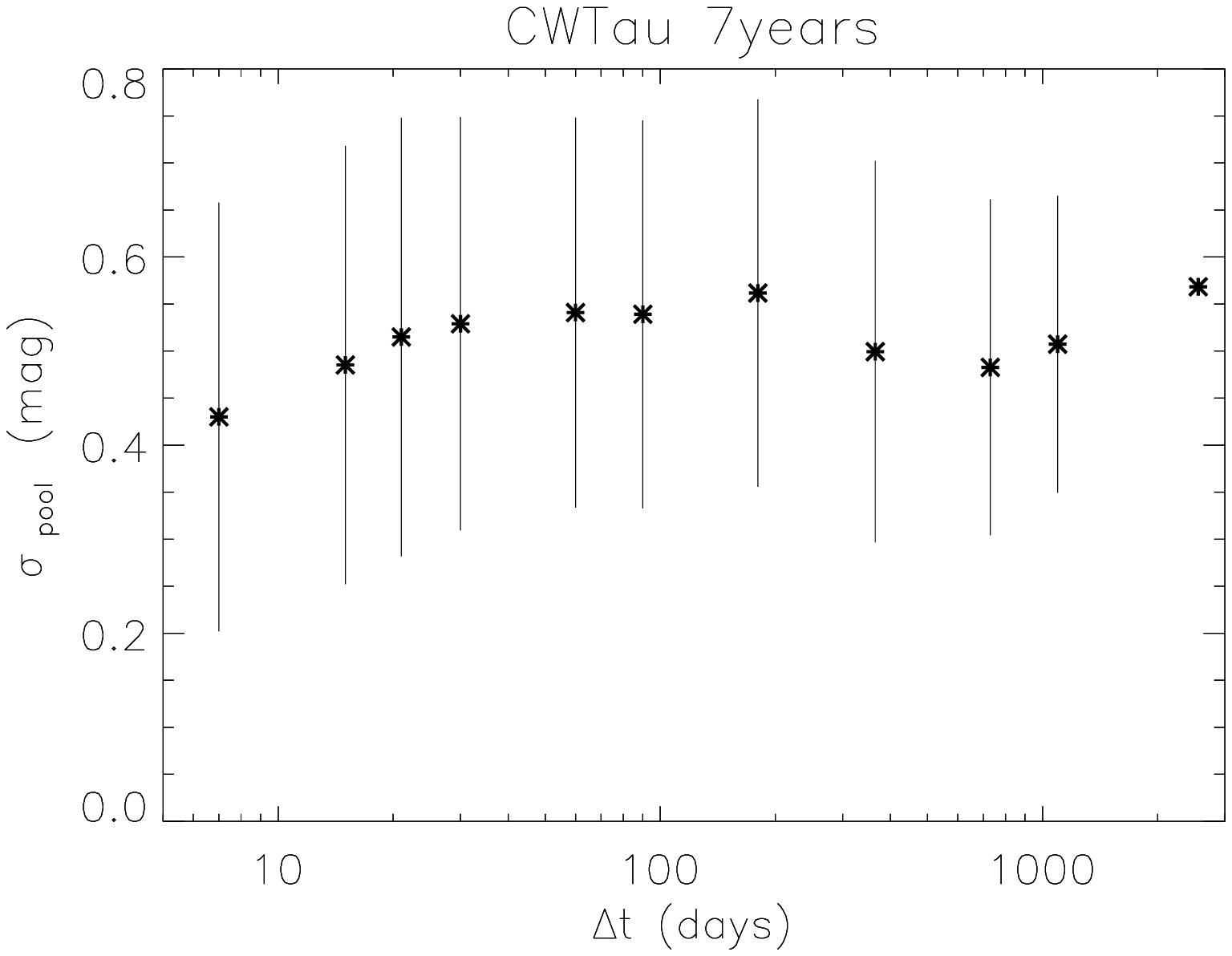}
\includegraphics[scale=0.45]{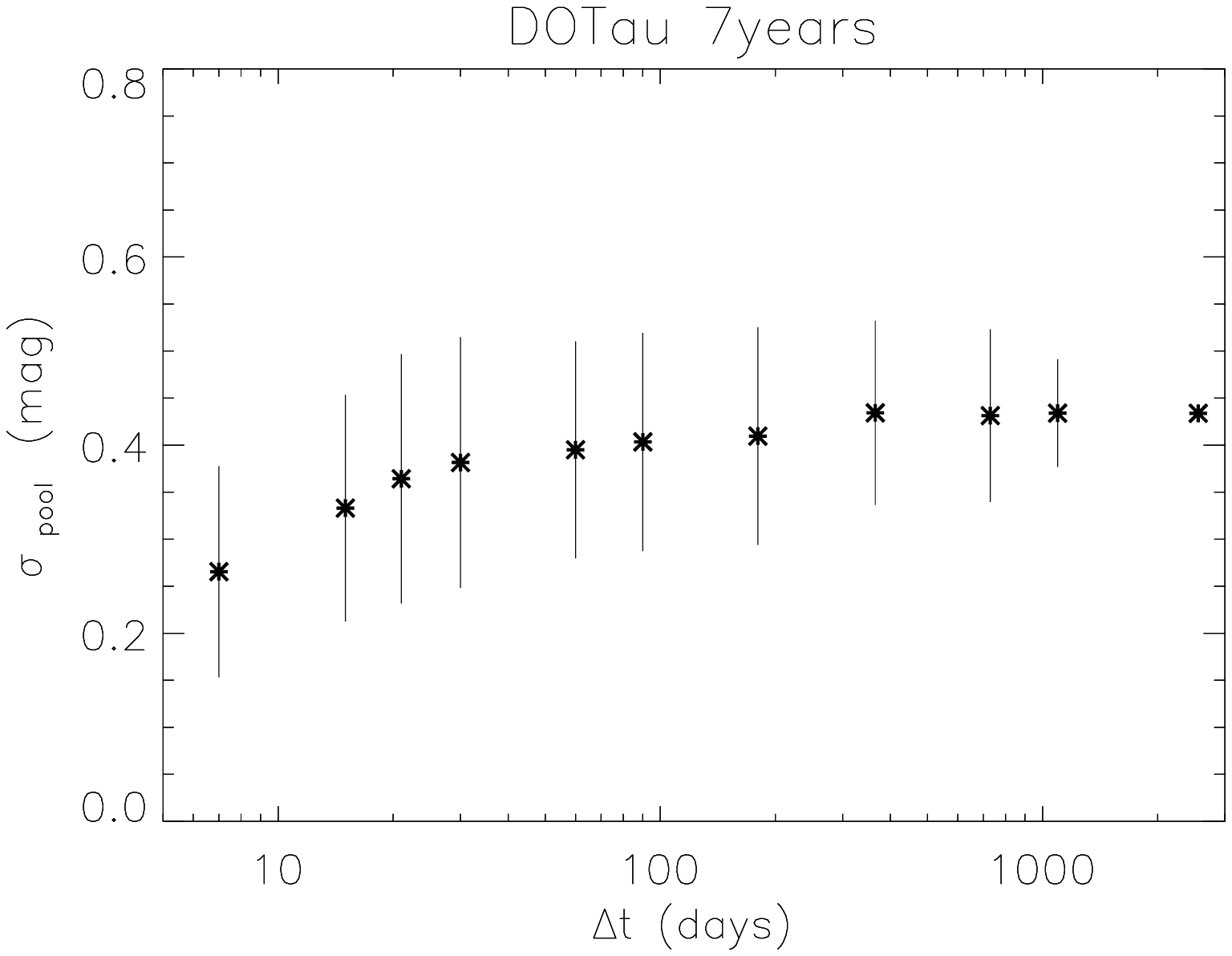}
\includegraphics[scale=0.45]{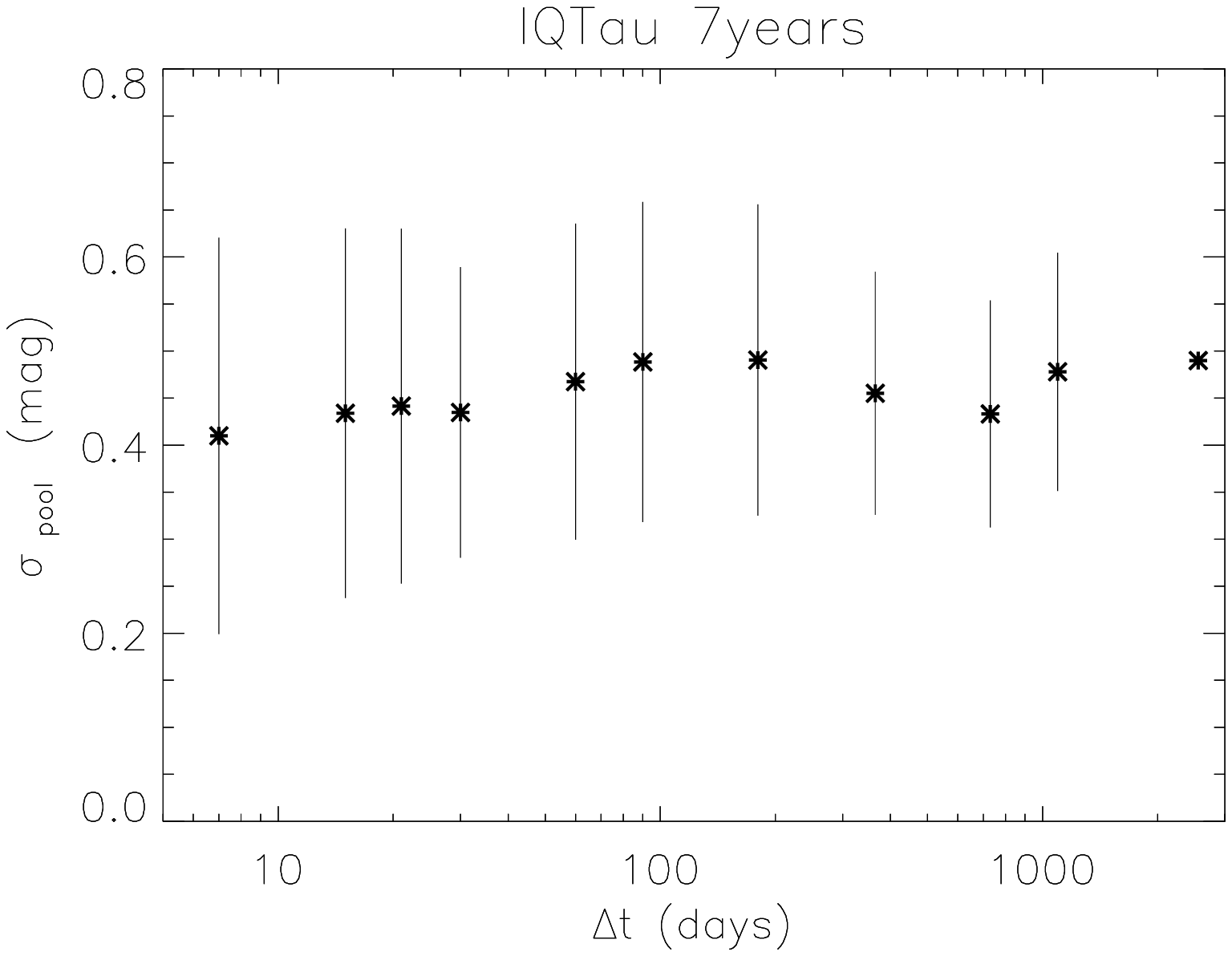}
\includegraphics[scale=0.45]{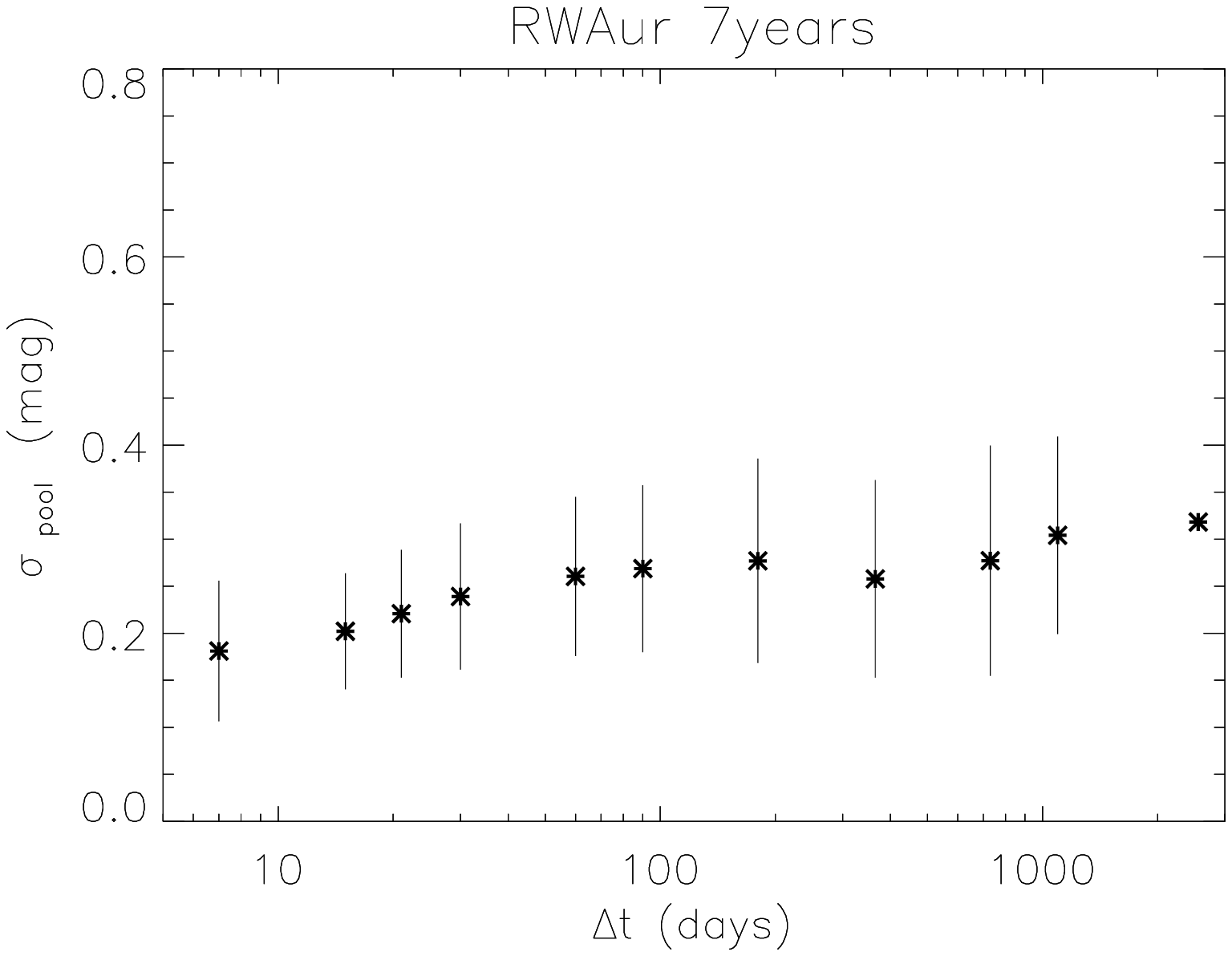}
\includegraphics[scale=0.45]{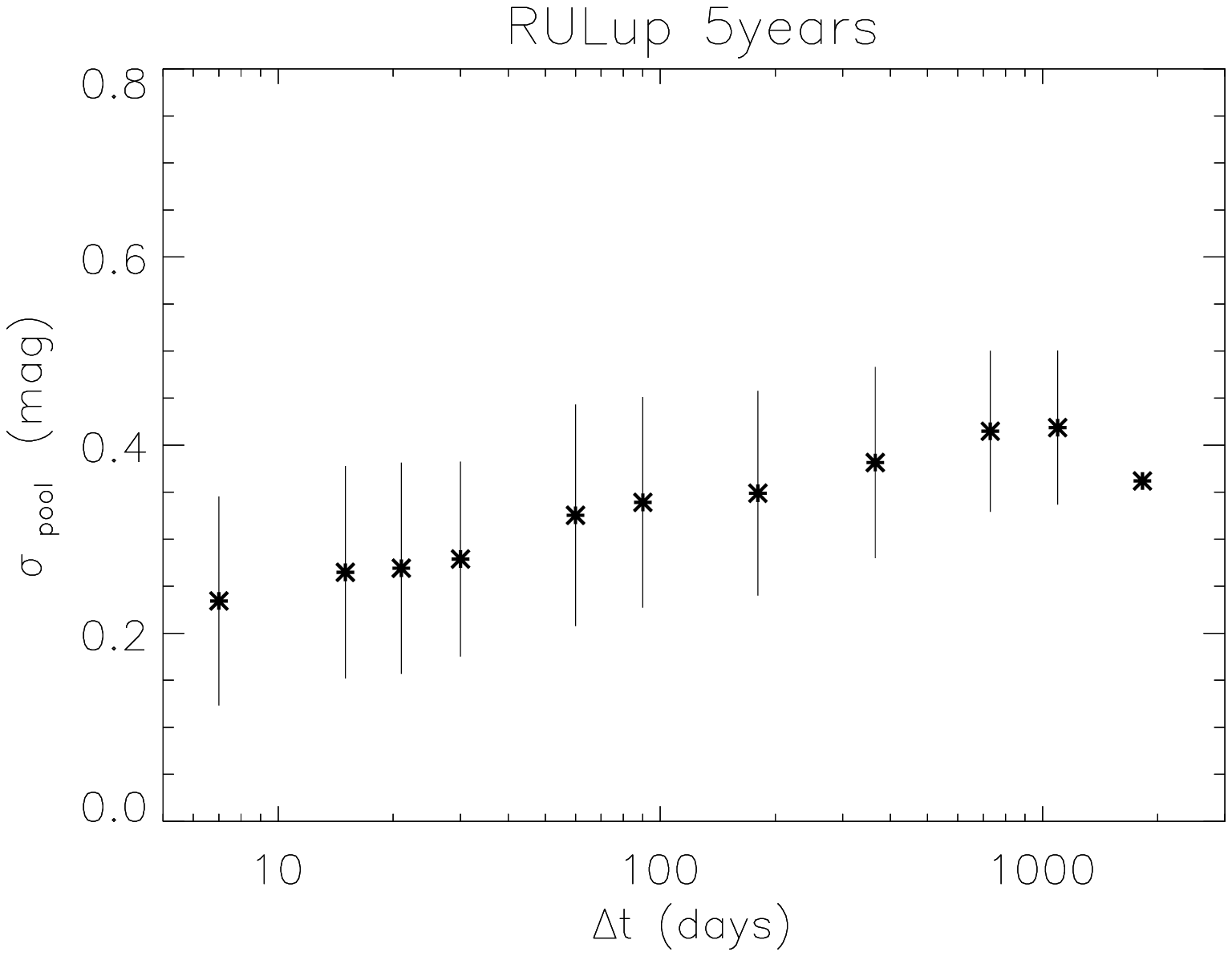}
\includegraphics[scale=0.45]{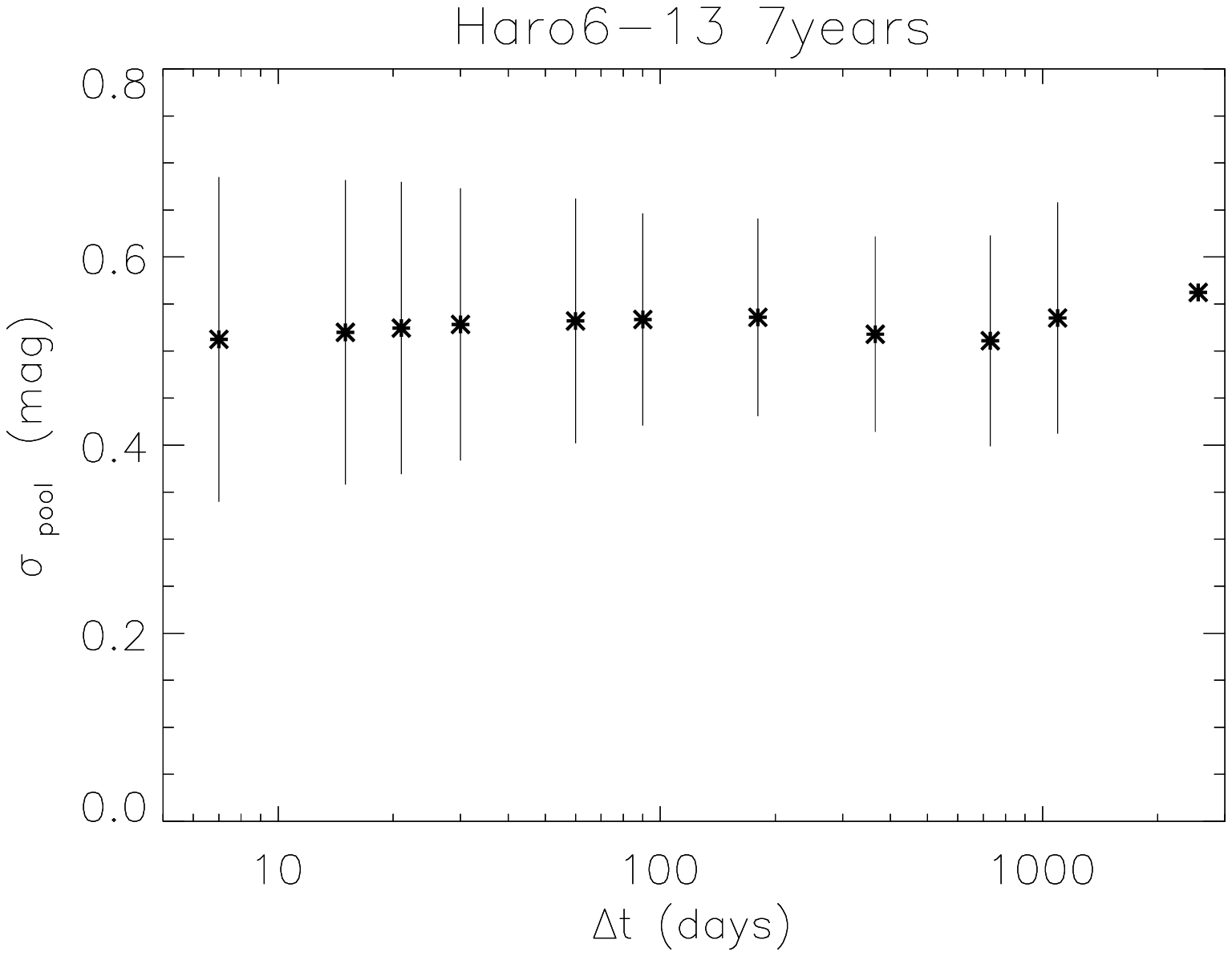}
\caption{Pooled sigma vs. timescale for 8 stars from our primary sample with strong variability}
\label{pool-quantile}
\end{figure*}

Eleven bins were pre-defined, with $\Delta t$ of a week, a fortnight, three weeks, one month, two months, three months, six months, a year, two years, three years and a decade. 
The decade subset, which corresponds to the entire sample, may differ from star to star, depending on the available data, but covers usually about seven years. One reason for choosing these timescales was to reflect the 'natural' sampling of the WASP lightcurves, which have large gaps between seasons. The pooled sigma was derived by moving the bins over the lightcurve in time steps of 1 day. Only bins with more than 50 data points were used in eq. \ref{pooledsigma_eq}, in order to avoid datasets with poor statistics. The pooled sigmas were then plotted and analysed as a function of $\Delta t$, see Fig.\ref{pool-quantile} for examples. 

The scatter of the pooled sigma for a given $\Delta t$ (errorbars in Fig. \ref{pool-quantile}) was computed through an equation analogous to eq. \ref{dev}, where $x_i$ was substituted with $\sigma_i$:

\begin{equation}\label{scatterpool}
\sigma_{\sigma_{pool}}=\sqrt{\sum_{i=1}^{n_{bins}}\frac{(\sigma_i-\sigma_{pool})^2}{N-1}}
\end{equation}

For comparison, we also produced the same plots with fixed time bins (i.e. 52 bins with $\Delta t$ of one week per year), which yielded similar results as the moving bins. We also tested using the 95-5\% quantiles instead of standard deviations (i.e. the range of the data between 5\% and 95\% of the full amplitude). This again gave qualitatively very similar results as the pooled sigma, and confirms that the trends discussed in the following do not depend on the specific metric used to quantify the variability.

\subsection{Results}
\label{subsec:Results}

For each star, we examined the pooled sigma as a function of time in plots such as the ones shown in Fig. \ref{pool-quantile}. Slightly over half of the sample show a flat dependence of pooled sigma on time, i.e. no significant increase over time. For about 1/3 of the sample, the pooled sigma increases initially but plateaus at some stage. Finally only a few objects in the sample show a steady rise of pooled sigma over the timescales considered here.

\begin{figure}
\includegraphics[scale=0.45]{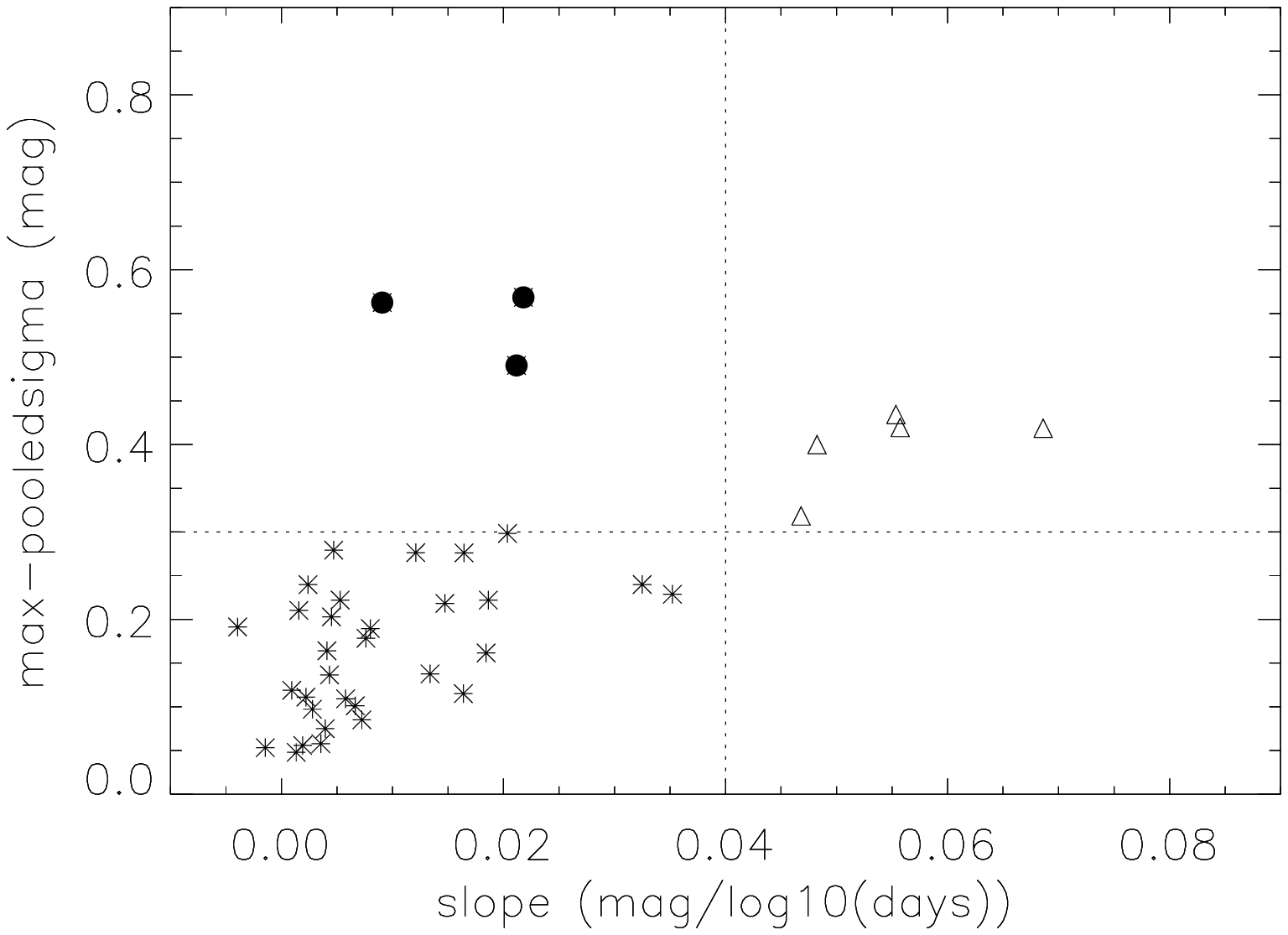}
\includegraphics[scale=0.45]{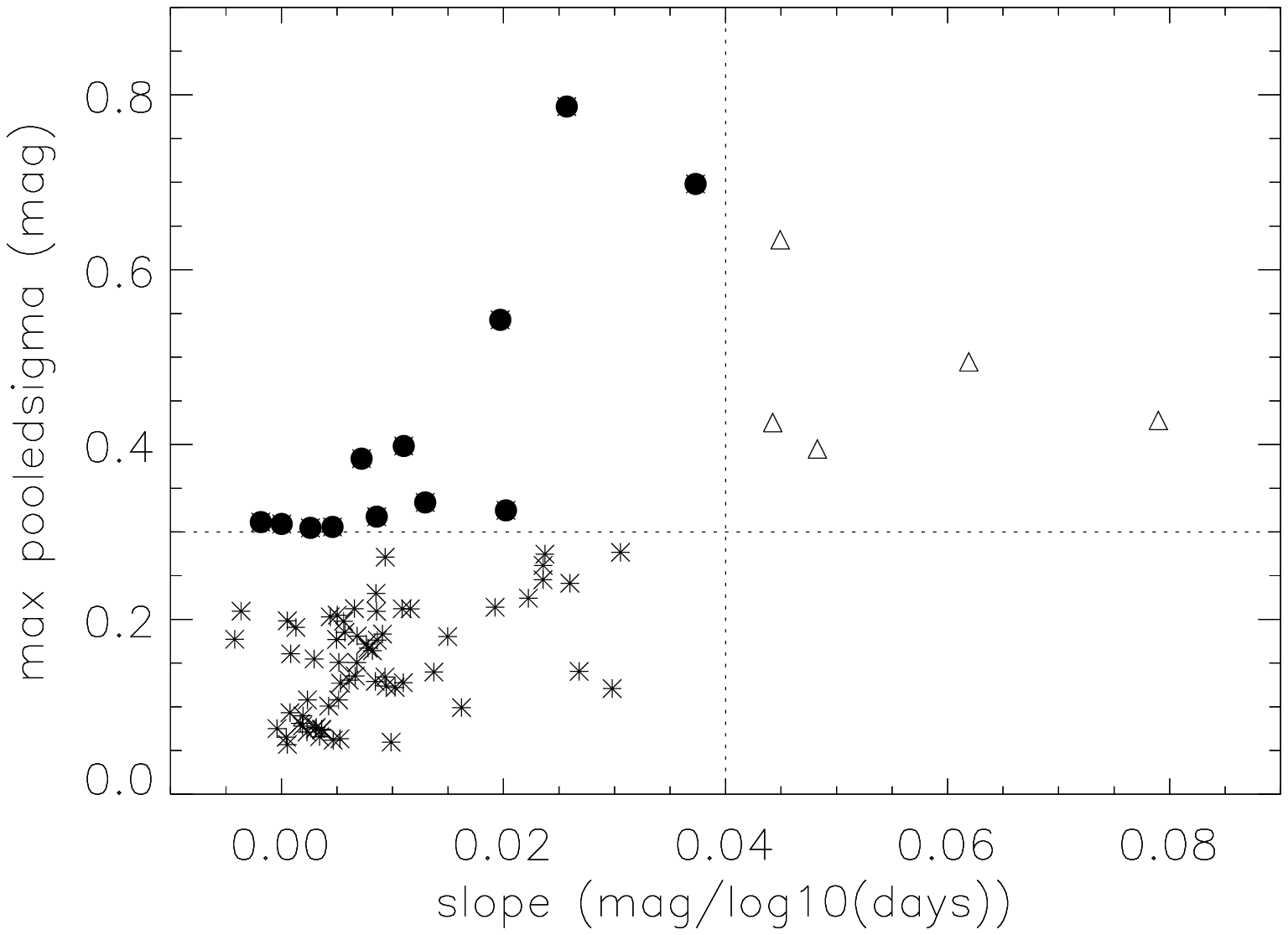}
\caption{Maximum pooled sigma vs. linear slope of the pooled sigma over time, for our primary sample (upper panel) and the control sample (lower panel).
Stars with significant positive slope ($>0.04$\,mag per week, group A) are marked with triangles; stars with maximum pooled sigma $>0.3$\,mag and slope $<0.04$\,mag are marked with filled circles (group B).}
\label{slopemaxvar}
\end{figure}

From the sequence of pooled sigma for each star we derived two quantities, the maximum pooled sigma $\sigma _{max}$ and the linear slope of the pooled sigma as a function of logarithmic time $S$ (which is in units of magnitude per $\log(d)$). The former provides an indication of the maximum variation of a star, the latter shows if and how much these variations evolve over time. In Fig. \ref{slopemaxvar} these two quantities are plotted against each other for our primary sample and for the control sample; they are listed in Table \ref{poolsig} for the primary sample.

The majority of the stars in both samples have $\sigma _{max}$ ranging from 0.05 to 0.3\,mag and a slope of $<0.04$\,mag per week. Considering the typical photometric errors (see Fig. \ref{mag-err}) this confirms that variability is ubiquitous in young stellar objects. However, only few objects show variability that causes $\sigma _{max}$ to exceed 0.3\,mag: 8/39 in the primary sample (21\%) and 17/81 (21\%) in the control sample. This is qualitatively consistent with previous findings derived by comparing two epochs of near-infrared photometry \citep{scholz2}: While variability in young stars is common, strong variability with amplitudes more than $\sim 30$\% is unusual, even for timescales up to several years.

For most stars, the slope of the pooled sigma $S$ is broadly consistent with zero. The mean of $S$ in our primary sample is 0.015 with a standard deviation of 0.017. Only five objects show clear evidence for a significant increase of the pooled sigma as a function of timescale (with $S>0.04$). For these objects, simple statistical tests ($\chi^2$ and t-test) confirm that the slope is not consistent with zero. The fraction of the objects with $S>0.04$ is 5/39 (13\%) in the primary sample and 5/81 (6\%) in the control sample. 

As a reminder, the control sample was selected to check whether or not our primary sample is biased towards highly variable stars. The comparison shows that this is not the case. The primary sample and the comparison sample exhibit similar variability characteristics. The fraction of objects with strong and/or long-term variations is around 20\% in both samples. Also, the distributions of slope $S$ and maximum pooled sigma are statistically indistinguishable. Both plots in Fig. \ref{slopemaxvar} show the same trends. The range of these parameters as well as the mean are similar as well. Thus, we conclude that our primary sample is not significantly biased in its variability properties and is representative of young stars with disk ('class II objects').

The scarcity of objects with significant long-term changes in the pooled sigma shows that the variability is typically dominated by the shortest timescales that we consider ($\sim$ weeks). This means that for {\it typical} T Tauri stars a few epochs of photometry covering about a month are fully sufficient to characterise the total extent of variability for timescales up to a decade. Similar results were found recently by other programs \citep{costigan,venuti}.

For objects that do show long-term changes, the plots shown in Fig. \ref{pool-quantile} reveal in some cases the typical timescale of variations as the point where the datapoints begin to 'plateau'. This is most obvious for AATau ($\sim 10$ weeks) and DOTau ($\sim 4$ weeks). These timescales can be compared to the results of the period search (see Sect. \ref{periodsearch}). All objects with $S>0.04$ also have $\sigma _{max}>0.3$, suggesting that long-term variability is driven by other mechanisms than the lower level variability on weekly timescales. An interesting feature in both samples is the fact that the stars with the highest maximum pooled sigma do not show long-term changes.

In the subsequent figures we will highlight objects with high $\sigma_{max}$ and high $S$. Stars with $S>0.04$ (AATau, CQTau,  DOTau, RULup and RWAur in the primary sample) will from now on referred to as group A (shown as triangles in Fig. \ref{slopemaxvar} and subsequent figures). Objects with $\sigma_{max} >0.3$ and $S<0.04$ (CWTau, IQTau, Haro6-13 in the primary sample) will additionally be referred to as group B (filled circles in Fig. \ref{slopemaxvar} and subsequent figures). For the further analysis we will mostly focus on group A, the stars with evidence for long-term variations significantly exceeding rotational timescales.

\begin{table}
\center
\caption{Results from the pooled sigma analysis for each star in the primary sample. The maximum pooled sigma and the slope of the linear fit to the pooled sigma $S$ are in the last two columns. The second column indicates if the star belongs to group A or B (see Sect. \ref{subsec:PooledSigma}).}
\begin{tabular} {@{}lccc@{}}
\hline
Star & Group &  $\sigma_{max}$ & $S$ \\
\hline
AATau             & A & 0.42 &  0.055   \\  
BPTau             & - & 0.14 &  0.013   \\
CITau             & - & 0.16 &  0.018   \\
CQTau             & A & 0.40 &  0.048   \\
CWTau             & B & 0.56 &  0.021   \\
DFTau             & - & 0.28 &  0.016   \\
DNTau             & - & 0.11 &  0.006   \\
DoAr24E           & - & 0.16 &  0.004   \\
DOTau             & A & 0.43 &  0.055   \\
DRTau             & - & 0.30 &  0.020   \\
FSTau             & - & 0.19 &  0.008   \\
FTTau             & - & 0.22 &  0.005   \\
GGTau             & - & 0.12 &  0.016   \\
GOTau             & - & 0.24 &  0.002   \\
GQLup             & - & 0.12 &  0.001   \\
Haro1-16          & - & 0.10 &  0.003   \\
Haro6-13          & B & 0.56 &  0.009   \\
Hen3-600A         & - & 0.06 &  0.002    \\
HKTauB            & - & 0.28 &  0.005    \\
HLTau             & - & 0.24 &  0.032    \\
HTLup             & - & 0.05 & -0.001    \\
HVTauC            & - & 0.14 &  0.004    \\
IMLup             & - & 0.07 &  0.004    \\
IQTau             & B & 0.49 &  0.021    \\
IRAS04189         & - & 0.18 &  0.008    \\
LkHa326           & - & 0.22 &  0.018    \\
LkHa327           & - & 0.22 &  0.015      \\ 
RULup             & A & 0.42 &  0.068     \\ 
RWAur             & A & 0.32 &  0.046      \\
RYTau             & - & 0.23 &  0.035  \\
TTauN             & - & 0.09 &  0.007  \\
TWA07             & - & 0.19 & -0.004  \\
TWHya             & - & 0.11 &  0.002  \\
UScoJ1604         & - & 0.28 &  0.012   \\
UZTauE            & - & 0.20 &  0.004  \\ 
V1121Oph          & - & 0.06 &  0.004   \\
V1149Sco          & - & 0.05 &  0.001   \\
V853Oph           & - & 0.21 &  0.001   \\
WaOph6            & - & 0.10 &  0.007       \\   
\end{tabular}                                                                                                          
\label{poolsig}                                                                                                        
\end{table}

\subsection{Period search}
\label{periodsearch}

We aimed to find a periodicity or quasi-periodicity in the lightcurves of stars in group A. Visual inspection of our lightcurves indicates that unique periods are usually not to be found. Several periods appear to be superimposed on each other, and in many cases periodic variations (or part of them) are visible only in some part of the lightcurves. Four different methods where applied in order to extract a trend in the lightcurves:

\begin{enumerate}
\item Lomb-Scargle periodogram \citep{scargle},
\item Epoch folding technique \citep{larsson},
\item Z-transformed Correlation Function (ZDCF) \citep{tal},
\item Phase Dispersion Minimization (PDM) \citep{stellingwerf}.
\end{enumerate}

Each of these techniques was tested, both on the entire lightcurve and on single 'chunks' of data corresponding to different years. This provided a set of candidate periods. Each of these periods was then tested by folding the lightcurve and examining it visually. Particular attention was paid when different methods yielded a similar period. In some cases the folded lightcurve showed underlying shorter periods, which were then tested again. This testing was applied several times, both on the entire data sample and on single chunks of time. The analysis was performed only on stars belonging to group A.

Two of four methods listed above turned out to be more useful in the period determination: the Epoch Folding and the ZDCF. The former tests different trial periods and folds the data onto them. It then returns, for each trial period, the $\chi^2$ computed to test the null hypothesis that the mean magnitude is the same over the entire trial period. A periodicity is given by a high $\chi^2$, because in this case the null hypothesis is not true and data points show instead a well defined pattern once the lightcurve is folded onto a specific period \citep{davies}. The latter method is based on the autocorrelation function, but it bins data according to the number of data instead of the time interval \citep{tal}. These two methods do not make assumptions about a specific shape of the periodicity, which makes them suitable for the partly irregular lightcurves in our sample. 

The most interesting outcome of this analysis is the presence of brightness variations with periods significantly longer than the rotational timescales, notably for AA Tau, CQ Tau, DO Tau, and RWAur, i.e. four out of five stars with long-term variability. In these cases the periods found in the WASP lightcurves range from $\sim 20$ to 60 days. These periods persist for at least several months and up to several years in the case of RWAur. To our knowledge, these long-term periodic changes have not been reported in the literature yet. We note that \citet{percy3} finds periods of similar duration for a few stars in our sample, but not the four periods reported here. For comparison, typical rotation periods for young stars at the ages of our targets range from 1 to 10\,d \citep{herbst07,xiao}. There is a tail of stars with longer rotation periods, but periods $>15\,d$ are considered to be very rare. For the 5 stars with long-term cycles, rotation periods of 5-15 days are reported in the literature. Therefore, the periods measured here cannot be attributed to the stellar rotation period.

\begin{figure}
\includegraphics[scale=0.5]{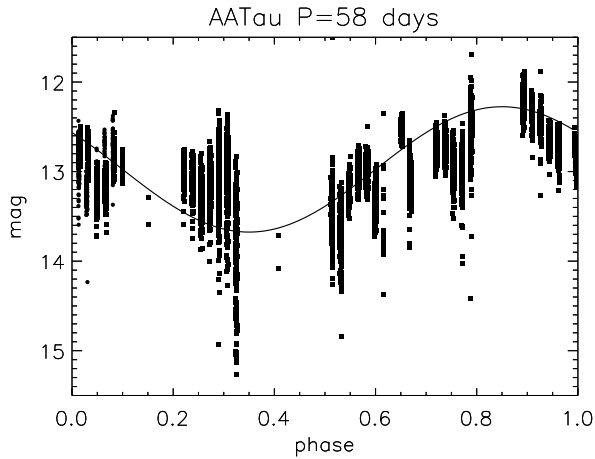}
\caption{Folded lightcurve of AATau between 24th September and 23th November 2011.}
\label{foldedAA}
\end{figure}

\begin{figure}
\includegraphics[scale=0.5]{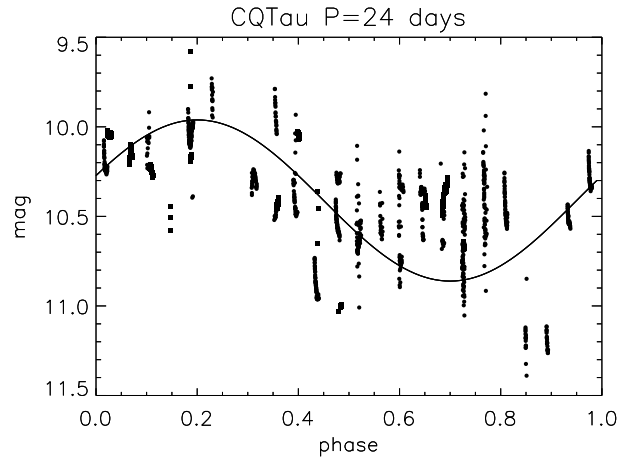}
\caption{Folded lightcurve for CQTau between 14th December 2006 and 24th February 2007.}
\label{foldedCQ}
\end{figure}

\begin{figure}
\includegraphics[scale=0.5]{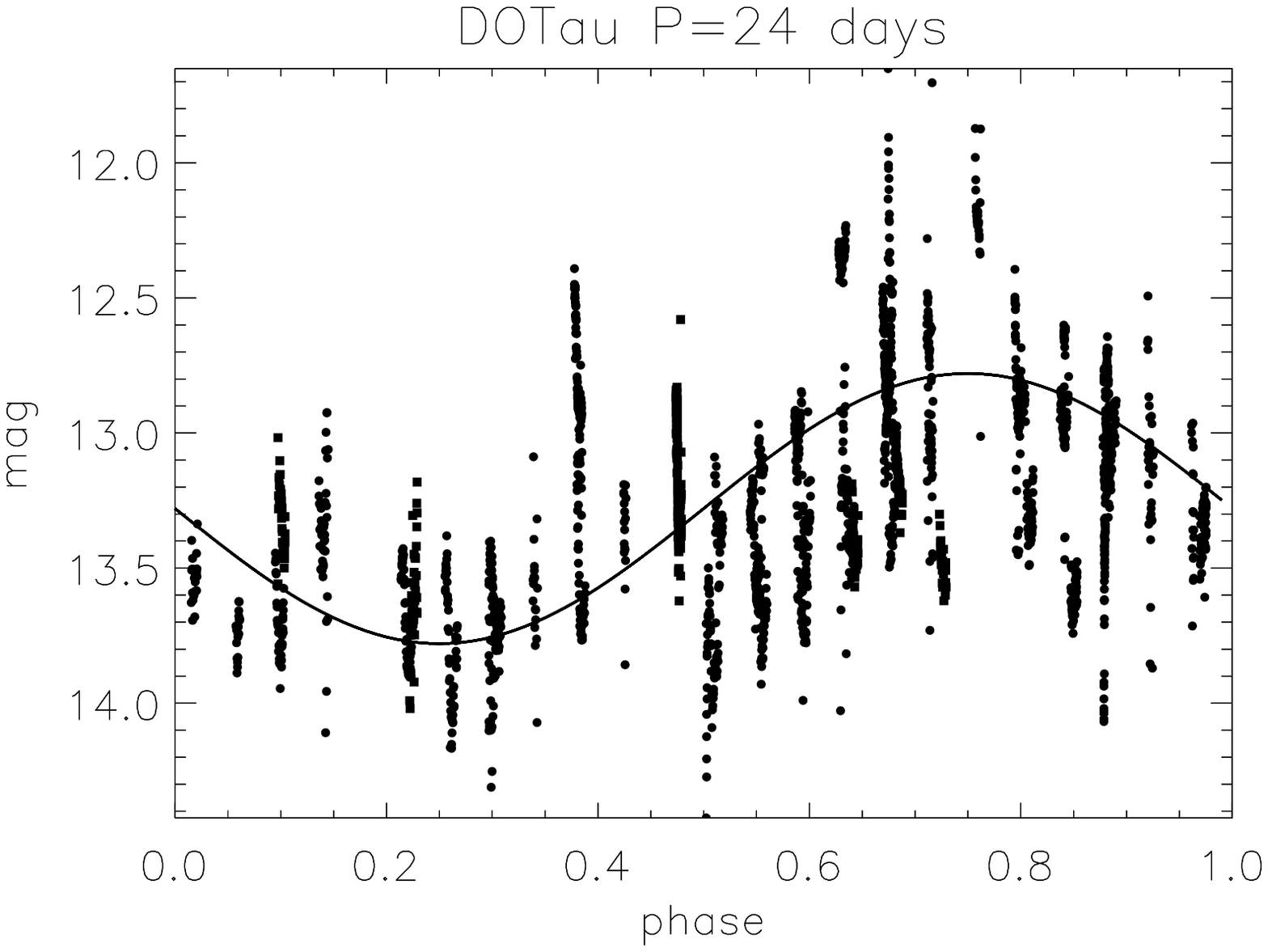}
\includegraphics[scale=0.5]{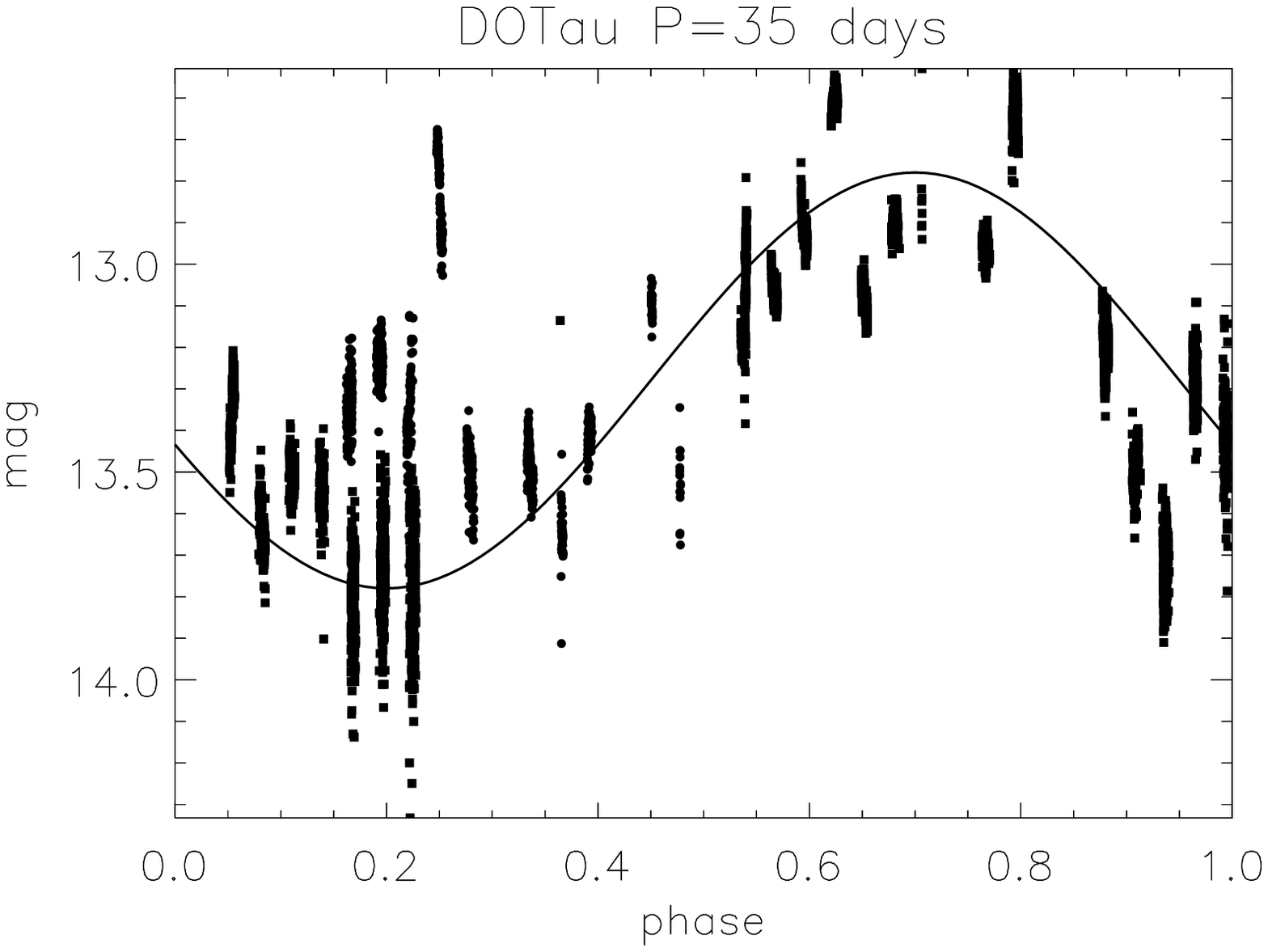}
\caption{Folded lightcurve for DOTau between 30th September 2006 and 24th January 2007 (upper panel) and between 7nth October 2011 and 25th November 2011 (lower panel).}
\label{foldedDO24}
\end{figure}

\begin{figure}
\includegraphics[scale=0.5]{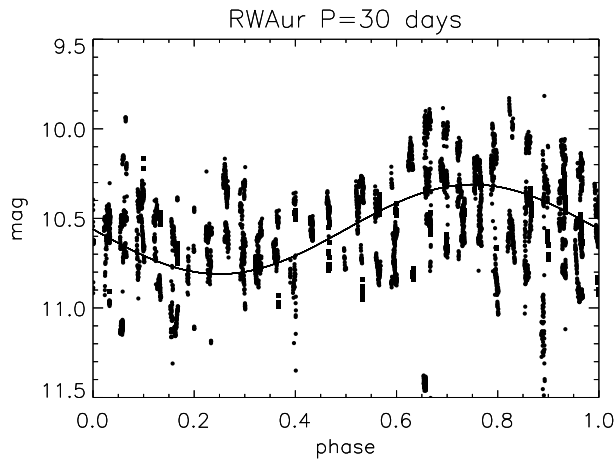}
\caption{Folded lightcurve for RWAur between 2nd August 2004 and 17th February 2011.}
\label{foldedrwaur}
\end{figure}

The periods confirmed by the algorithms and by visual inspection are discussed further below. For the most pronounced periods we show the (partial) lightcurves folded to the best period in figures \ref{foldedAA} to \ref{foldedrwaur}.

{\bf AATau:} The lightcurve has data at irregular intervals from 2004 to 2011. The most interesting feature is the part of the lightcurve between September and November 2011 (Fig. \ref{foldedAA}), which can be fitted with a sine curve having a period of 58 days. Observations in earlier years do not show this behaviour. We also note that the pooled sigma for this star saturates after around 10\,weeks, a timescale comparable to the measured periodicity (see Fig. \ref{pool-quantile}). The ample scatter is due to shorter variations, with a period around 5 days, which are slightly shorter than the typical rotation period of this star: 8.4 days according to \cite{bouvier} or 8.2 according to \cite{artemenko}. 

{\bf CQTau:} This star was observed from 2004 to 2010, with the majority of the observations being between October 2006 and February 2007. At the end of this time interval a period of 24 days was found (Fig.~\ref{foldedCQ}).

{\bf DOTau:} The star was observed at irregular intervals between 2004 and 2011. In year 2004, a 5-6 days period seemed to fit the data, which corresponds to half of the rotation period determined by~\cite{osterloh}. Between September 2006 and January 2007 a period of 24 days shows a good fit on the folded lightcurve (Fig.~\ref{foldedDO24}, upper panel). The scatter around the fitted sine curve is due to variations on a shorter timescale of 6-7 days, again shorter than the rotation period of 12.5 days found by \citet{osterloh}. Between August 2010 and February 2011 a tentative periodicity of 17 days was found. Between October and November 2011, instead, variations up to half a magnitude are visible with a period of 35 days (Fig.~\ref{foldedDO24}, lower panel). For comparison, the pooled sigma indicates the beginning of a plateau for a timescale around 4 weeks, which is in line with the long periods measured here.

{\bf RWAur:} This star was observed from 2004 to the beginning of 2011, with most of the observations being before the end of 2008. This is the only case where a single period of 30 days folded very well over the entire lightcurve (Fig. \ref{foldedrwaur}). This period is certainly greater than the rotation period, given as 5.6 days in the literature \citep{dodin}.

\section{Correlations with stellar and disk parameters}
\label{corr}

In order to investigate the origin of the long-term variations discussed in the previous section, we looked for correlations between several physical quantities versus the slope of the increase of pooled sigma over time, $S$. We are exploiting here the fact that our sample is well-studied and most of the parameters for stars and disks are known. The strength of correlations was tested by comparing the Pearson's correlation coefficient with critical values at 0.05 level of significance for a two-tailed test. The null hypothesis consists in having no correlations and it is rejected, and hence there is a significant correlation, when the absolute value of the correlation coefficient $r$ is greater or equal the critical value $r_c$ (taken from tables of critical values for Pearson correlation). The resulting correlation coefficients and the critical values are listed in Table \ref{corrcoeff}. We show all tested parameters in plots vs. $S$ in Fig. \ref{slopecorrplots} (with the exception of binarity, which is discussed separately).

\begin{table}
\center
\caption{Results from the correlation testing in Sect. \ref{corr}. We list the correlation coefficient for the plots
shown in Fig. \ref{slopecorrplots} together with the critical value. Significant correlations are marked in bold face.}
\begin{tabular} {@{}lcc@{}}
\hline
Quantity & $r$ & $r_c$  \\
\hline
Spectral type                                      & $-0.18$ & $0.33$ \\
Inclination                                        & $0.005$ & $0.42$ \\
Flux at 1.3\,mm (scaled to distance of Taurus)     & 0.27    & 0.37 \\ 
Spectral slope $\alpha_{\lambda}$ at 3.4-4.6$\mu$m & 0.51 & 0.33 \\
Spectral slope $\alpha_{\lambda}$ at 24-70$\mu$m   & -0.21 & 0.44 \\
Spectral slope $\alpha_{U-B}$                      & -0.73 & 0.60 \\
\hline
\end{tabular}                                                                                                          
\label{corrcoeff}                                                                                                        
\end{table}

\subsection{Binarity}

\begin{figure}
\includegraphics[scale=0.5]{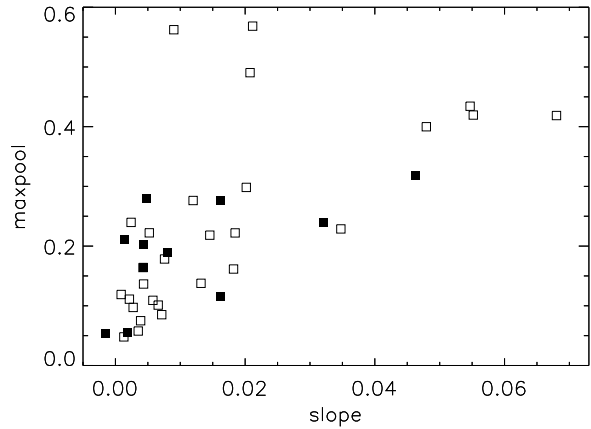}
\caption{Same plot as in Fig.~\ref{slopemaxvar}, where in open squares there are single stars, in filled squares binary stars.}
\label{binaries}
\end{figure}

We checked for known companions to stars in our sample in the literature. We focused on objects in orbits detectable by high spatial resolution imaging, with separations comparable or larger than $\sim 10$\,AU. In Table \ref{t1} we list the multiplicity and the references. The sample contains 12 known multiple systems, among them 3 triples. Fig. \ref{binaries} is identical to Fig. \ref{slopemaxvar}, upper panel, but the multiple systems are marked. The binaries are scattered over the diagram; there is no clear trend with $S$. We confirmed with an KS test that the distributions of $S$ of single stars and multiples are indistinguishable. Thus, the presence of wide companions does not seem to have an effect on the long-term variability.

\subsection{Spectral Type}

We searched for a correlation between $S$ and the spectral type of the star, which serves as a proxy for its effective temperature and also mass. There is no systematic difference between group A or B and the other stars; variability characteristics are comparable over a wide range of spectral types and hence stellar mass.

\begin{figure*}
\includegraphics[scale=0.45]{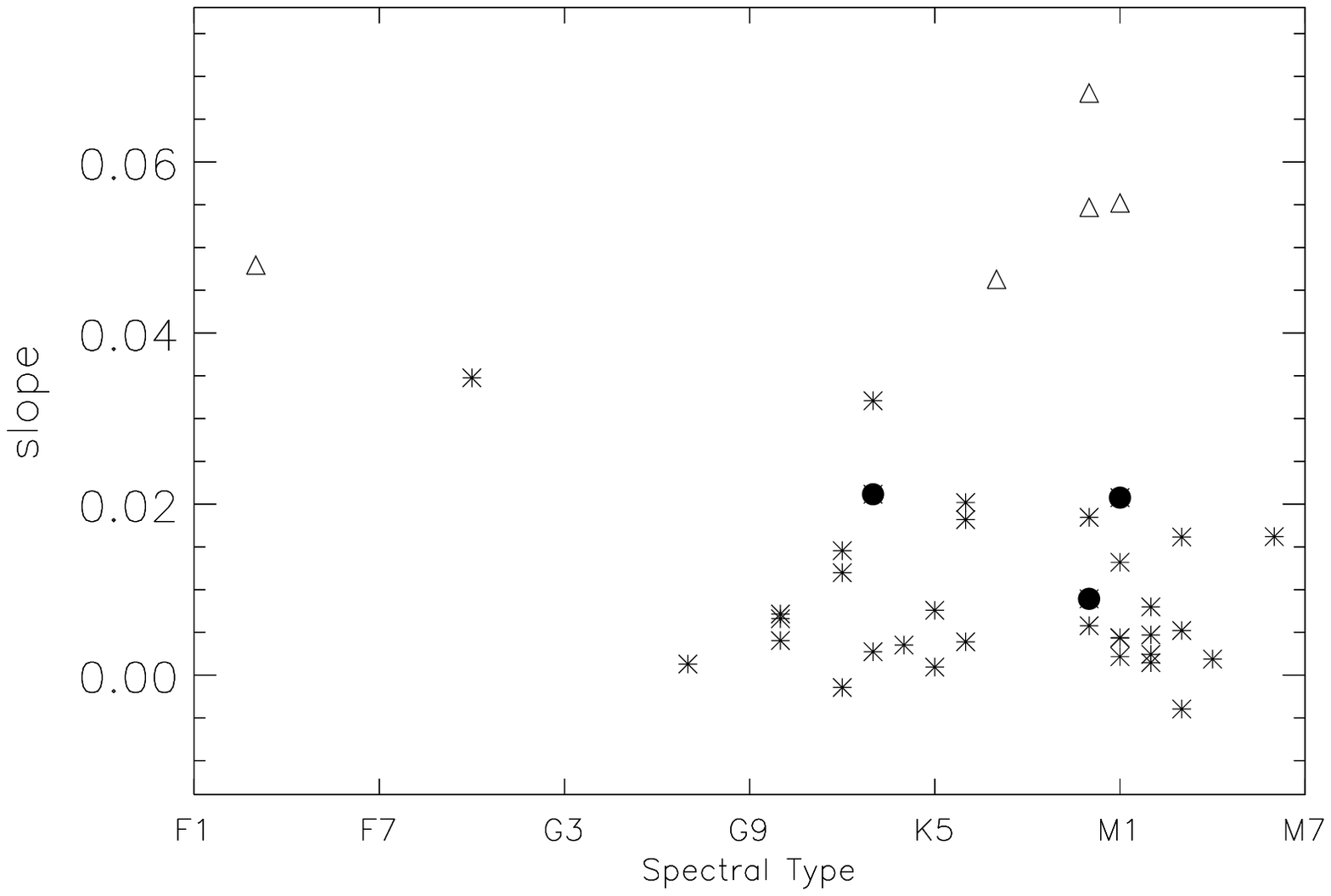}
\includegraphics[scale=0.45]{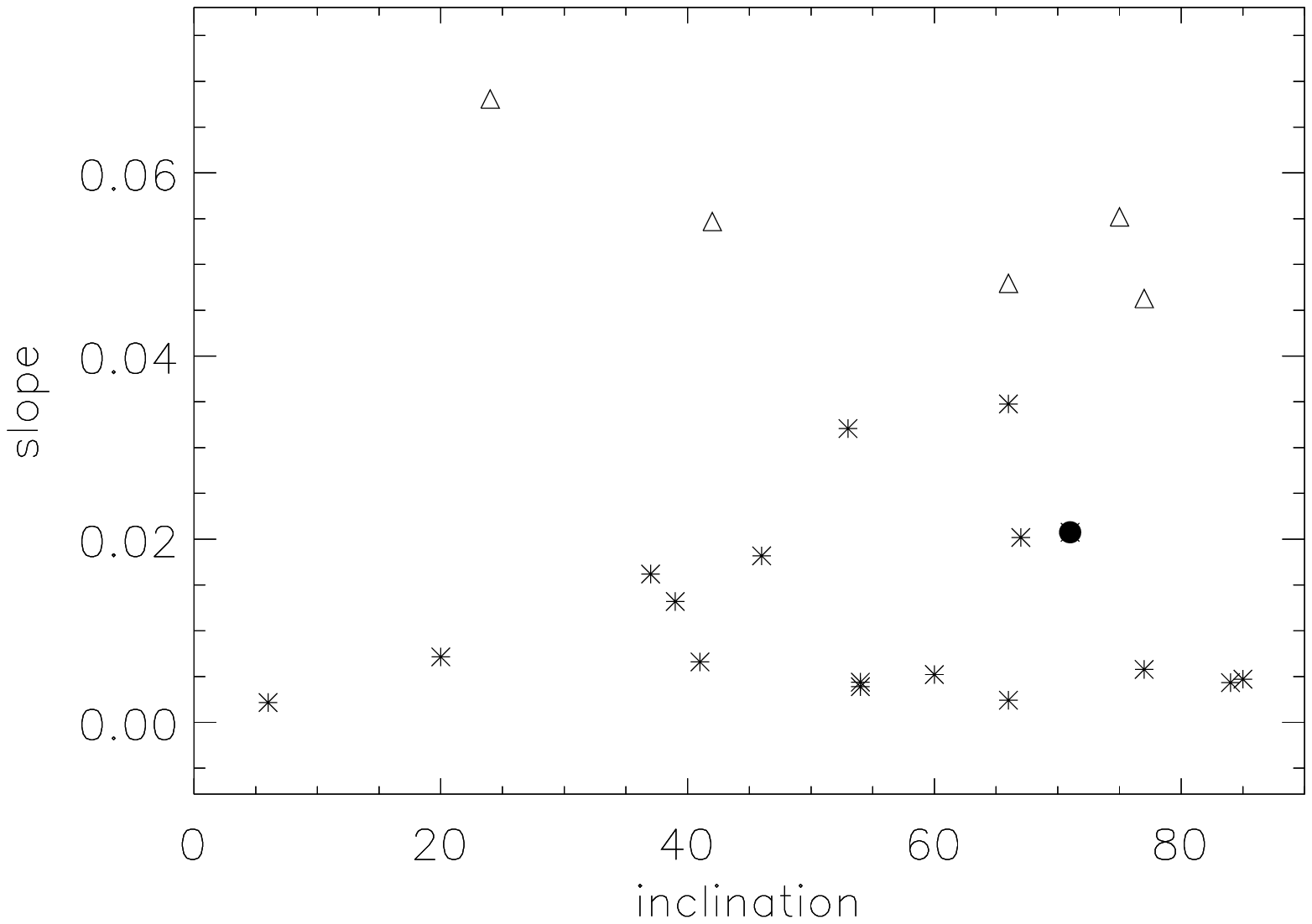}
\includegraphics[scale=0.45]{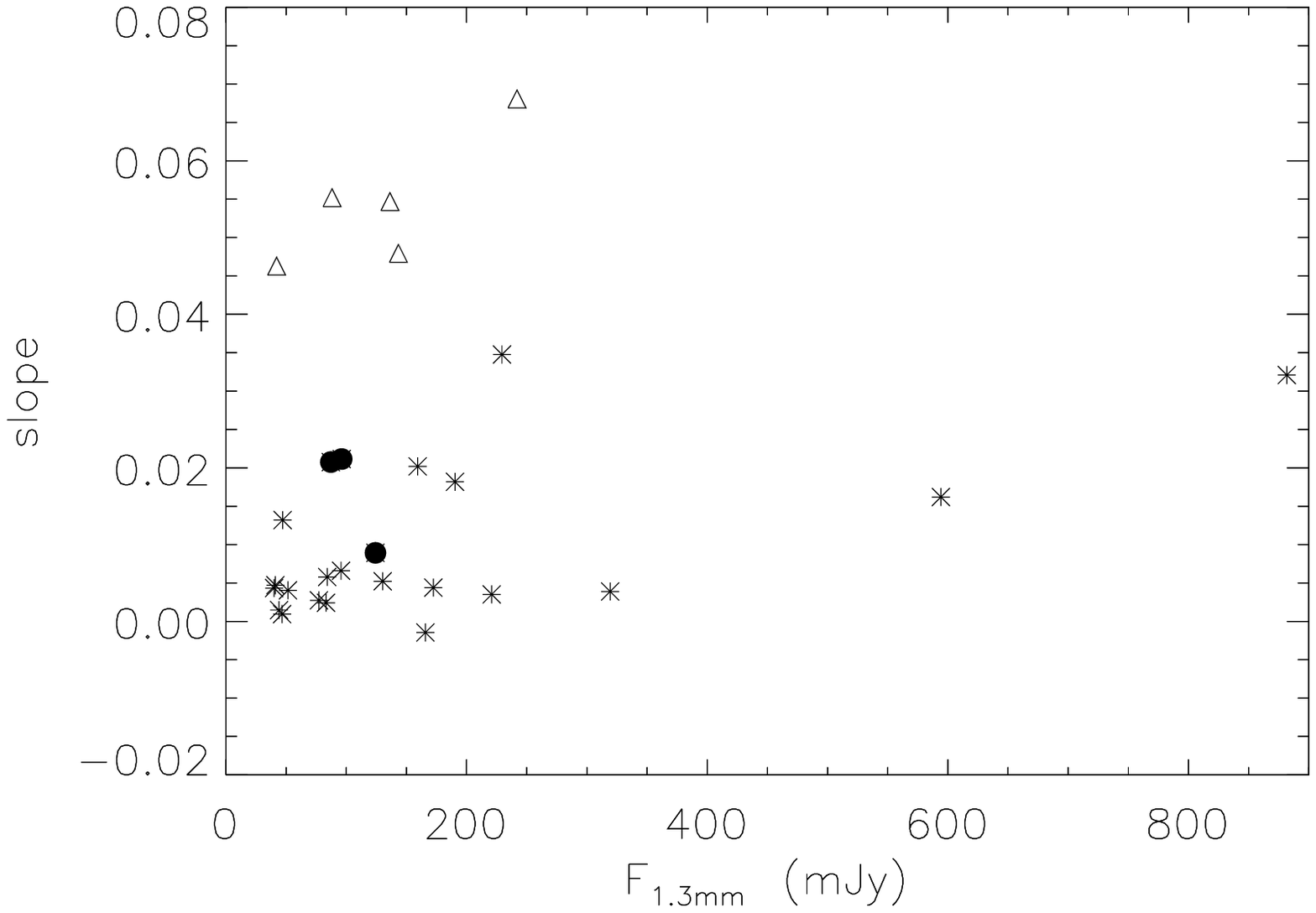}
\includegraphics[scale=0.45]{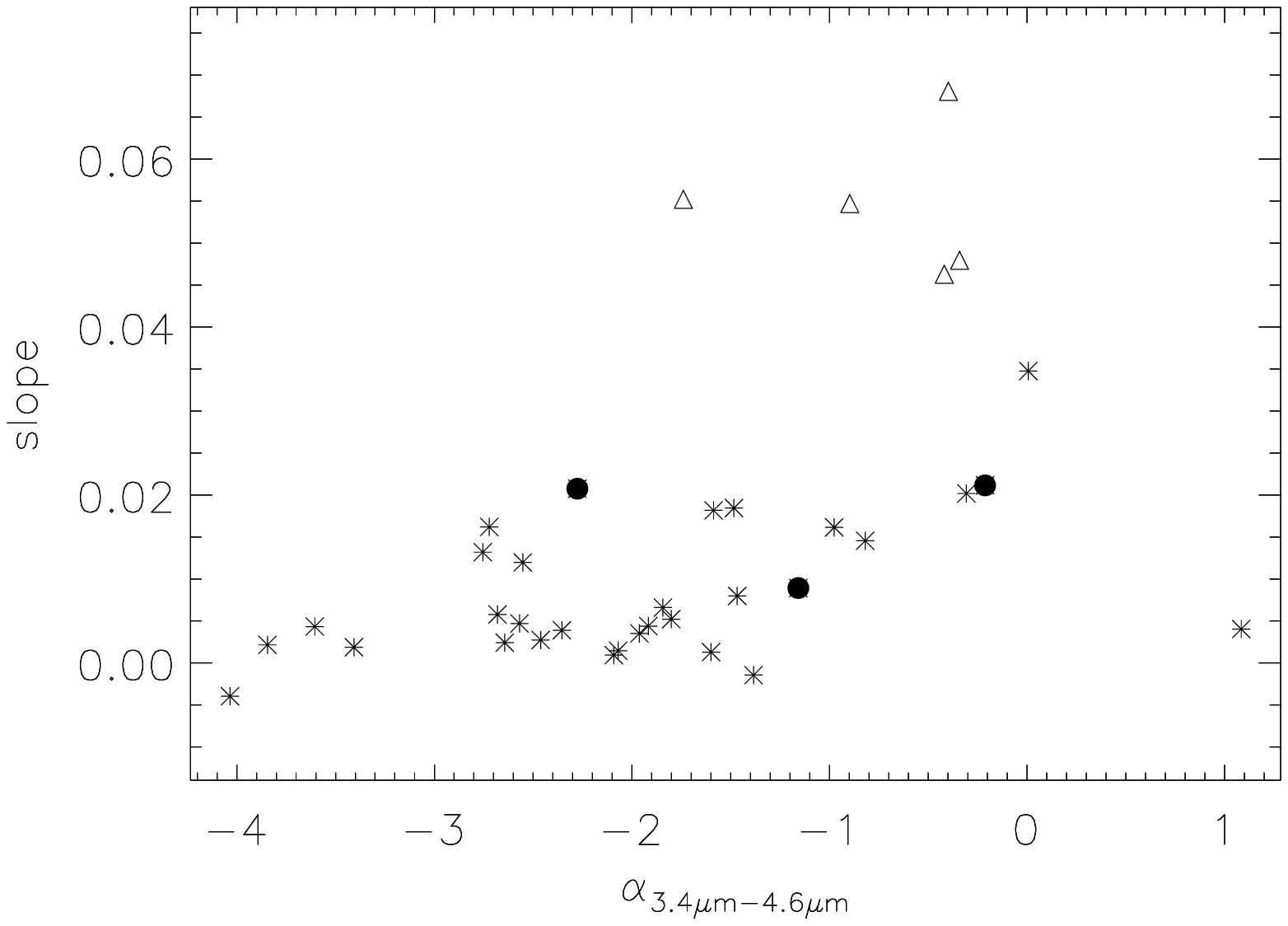}
\includegraphics[scale=0.45]{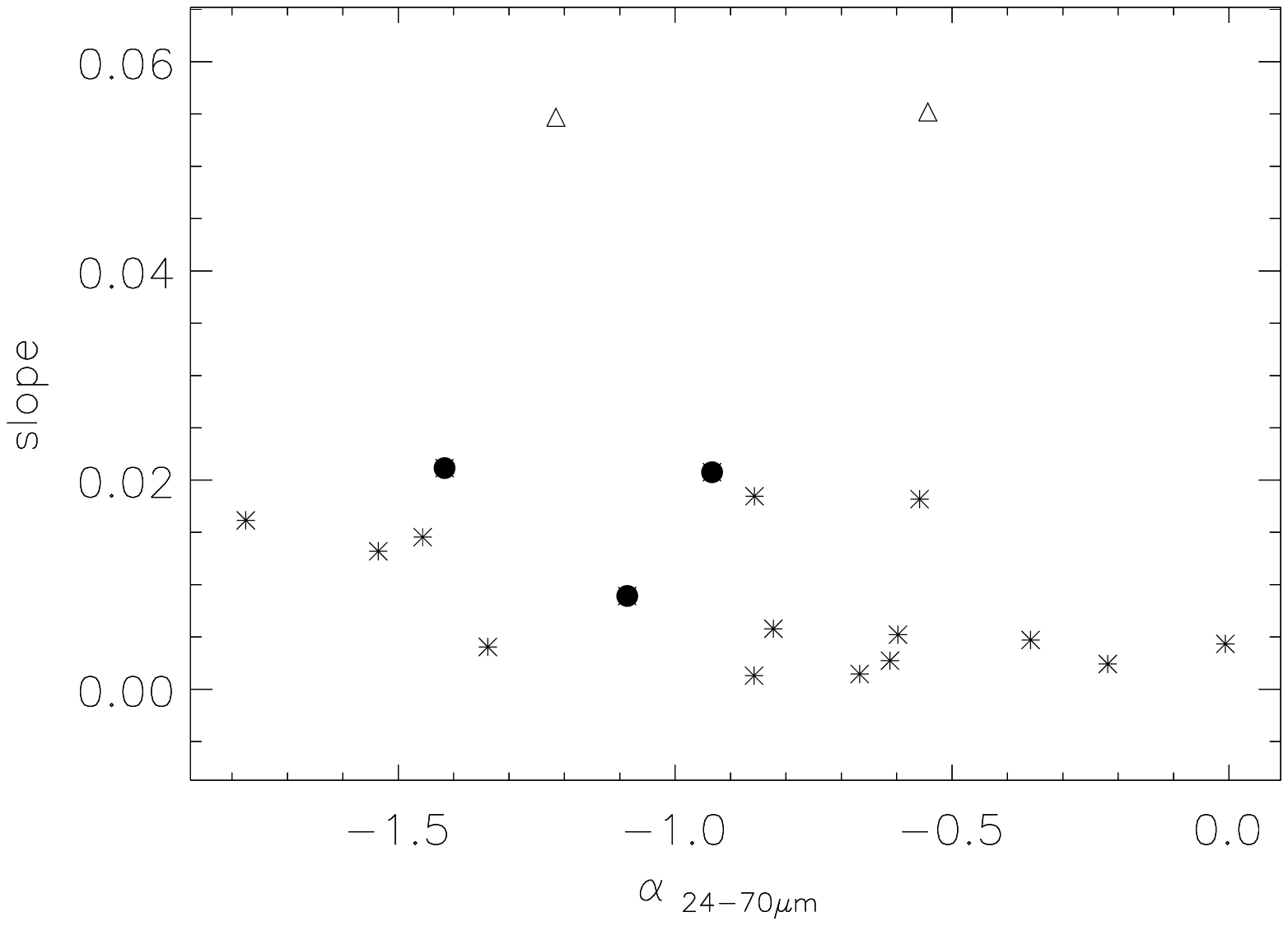}
\includegraphics[scale=0.45]{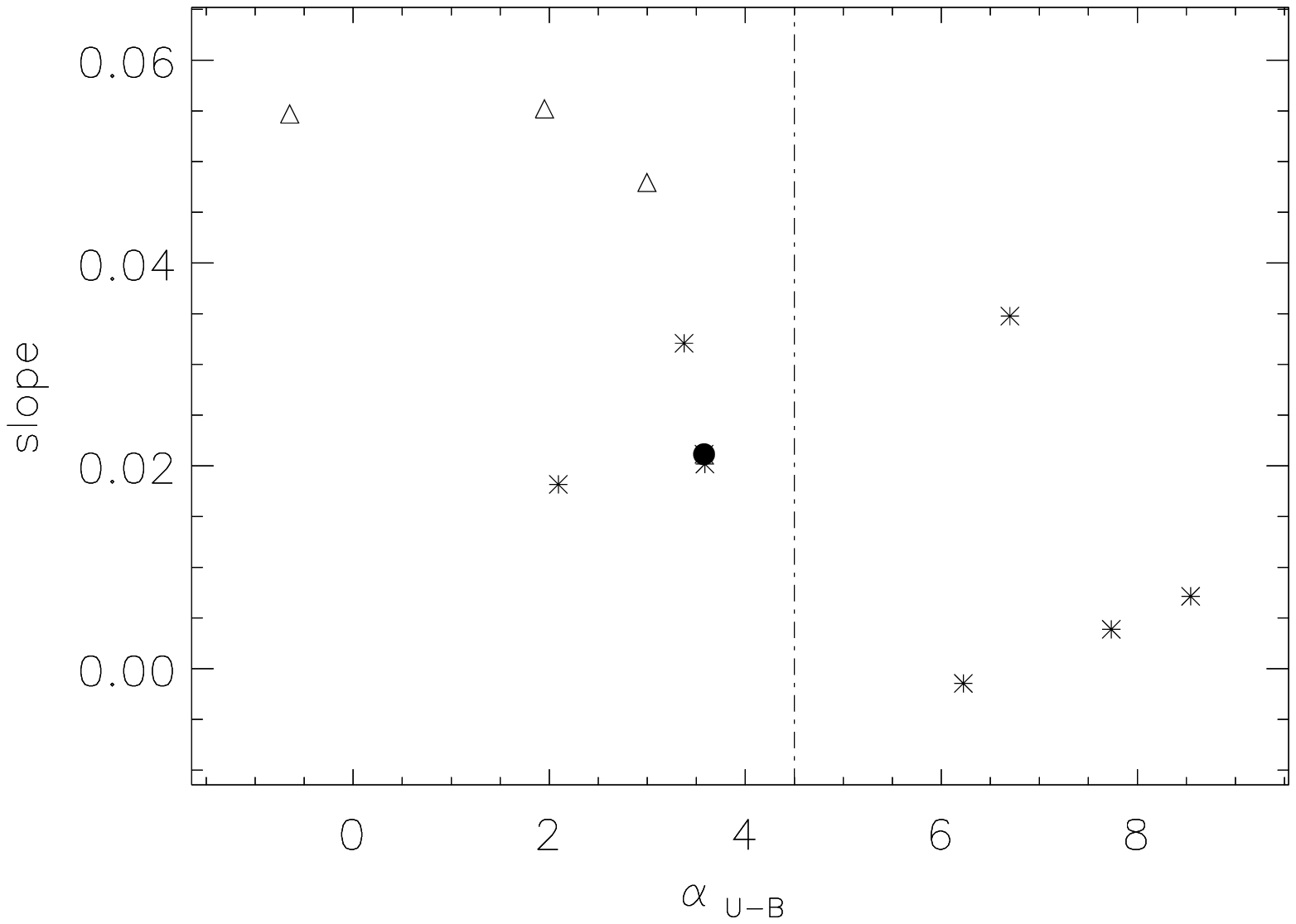}
\caption{The linear slope of the pooled sigma increase over time plotted vs. a sequence of stellar and disk parameters. In triangles stars from group A, in filled circles stars from group B. See the discussion in Sect. \ref{corr}}
\label{slopecorrplots}  
\end{figure*}

\subsection{Disk Inclination}
\label{subsec:diskinclination}

If the variability is due to variable extinction or obscuration caused by inhomogenities in the disk, we would expect to see a dependence with the disk inclination, in the sense that variability is more pronounced for disks seen close to edge on. Under high inclination, the line of sight has a higher likelihood of transsecting parts of the disk. For a large fraction of our sample (22/39), the disk inclination has been estimated in the past using a variety of techniques, including resolved submm/mm interferometry, high-resolution infrared imaging, variability modeling, modeling the unresolved SED, and comparisons of $v\sin{i}$ with rotation period. The uncertainty in these values depends on method but is typically in the range of 10-20\,deg. The inclinations for individual objects and corresponding references are listed in Table \ref{t1}.

For the full sample, no correlation is found between inclination and slope of the pooled sigma. In particular, stars with strong long-term variability are found at low and high inclinations. Thus, extinction or obscuration by the inner disk as the only explanation for long-term variability is not applicable.
From Fig. \ref{slopecorrplots}, it does appear that RU Lup and DO Tau, the only long-term variables with inclinations less than 60\,deg, are outliers. But the evidence for their relatively low inclinations is solid. Taking these objects out, there does seem to be a trend of increasing $S$ towards higher inclinations, which could mean that disk obscuration/extinction matters at least in some objects with long-term trends. We also note that the inclination for RWAur is debatable; we use the value derived from inner disk modeling (77\,deg), but for the overall disk a smaller angle is reported (46-60\,deg, \citet{cabrit}).

\subsection{Disk Mass}

The total disk mass, inferred from submm/mm observations, is one of the main global parameters for the disk. The bulk of the mass is in the outer disk and declines gradually with age. Most of our sample has been observed with single-dish instruments at 1.3\,mm wavelength \citep[e.g.][]{beckwith90,osterloh95,nuernberger97,nuernberger98,andre94}, a spectral range that traces the disk mass well: $M_{\nu}=\frac{F_{\nu}d^2}{\kappa_{\nu}B_{\nu}}$. Here, $F_{\nu}$ is the measured flux, $d$ is distance, $\kappa$ the dust opacity and $B_{\nu}$ the radiation from a blackbody with the temperature of the dust. We do not find any correlation or trend between 1.3\,mm flux and the slope of the pooled sigma $S$, after scaling all fluxes to the distance of Taurus (see Table \ref{fig6quant}). This is indicating that the long-term variability is unrelated to the global state of the disk. 

\subsection{Infrared emission}
\label{iremission}

Emission from the star is intercepted, absorbed and re-emitted by the dust grains in the disk. The regions of the disk close to the star (0.1-1\,AU) emit near- and mid-infrared radiation, while regions further out (1-100\,AU) are traced by far-infrared and submm radiation \citep{carmona2010}. To probe the relation between inner disk properties and variability, we collected infrared fluxes for all our targets and calculated spectral indices $\alpha_{\lambda}$:

\begin{equation}
\alpha_{\lambda}=\frac{\lambda_1\log_{10}F_1-\lambda_2\log_{10}F_2}{\lambda_1-\lambda_2}
\end{equation}

We use fluxes from the WISE mission \citep{wise} to derive $\alpha_{\lambda}$ between 3.4$\mu$m and 4.6$\mu$m. At longer wavelengths $\alpha_{\lambda}$ was computed between 24$\mu$m and 70$\mu$m, using Spitzer/MIPS fluxes \citep{luhman,infrared3}. The results are listed in Table \ref{fig6quant}.

In Fig. \ref{slopecorrplots} we show $\alpha$ for 3.4-4.6$\,\mu m$ for our sample vs. the slope $S$. The statistics is strongly affected by three outliers: IRAS4189+2650, HLTau and TTauN, which have $\alpha_{\lambda}>3$ and are not in the plot. In all three cases, the WISE fluxes are significantly contaminated by emission from nearby sources and nebulae. After the exclusion of the three outliers, the critical value becomes $r_c=0.33$, and the correlation coefficient is 0.51, demonstrating a significant correlation between long-term variability and near-infrared slope. All stars in group A and B have spectral index greater than $-2.3$. This analysis suggests that long-term variability is related to the properties of the inner disk. A large spectral slope in the mid-infrared is the hallmark of a dusty disk with large scaleheight close to its inner edge, either due to a wall or due to flaring. 
On the other hand, for the slope between 24 and 70$\,\mu m$, the correlation coefficient in this case is clearly below the critical value. Stars in group A or B do not show any particular trend with the spectral index in this region. 

\subsection{Ultraviolet emission}

Accretion from the disk to the star produces excess emission in the ultraviolet and blue part of the spectrum. To probe this part of the spectrum, we collated UBV-band fluxes from the literature, specifically from \citet{uv6,uv7,uv8,uv9,uv10,ofek,Richmond,dewinter}. To correct for extinction, we calculated $A_J=R(J)\times E(B-V)$ where the colour excess $E(B-V)$ is defined as the difference between the observed and the intrinsic colours, $E(B-V)=(B-V)-(B-V)_0$. We take the intrinsic colours from \citet{pecaut} and the value of the constant $R(J)$ from \citet{yuan}. Once $A_J$ was known, each wavelength-specific extinction, $A_{\lambda}$, was derived through the $A_{\lambda}/A_J$ ratio values provided by \cite{mathis} in their Table 1. UBV-band magnitudes are available for 11 objects, with three in group A (AATau, CQTau, DOTau) and one in group B (CWTau). 

Spectral indices were computed for the extinction corrected colours $U-B$ (see Table \ref{fig6quant}). Although the sample is limited in size, we find a significant anti-correlation between $S$ and the $U-B$ spectral slope. Stars in group A and B are on average bluer than the full sample, with $\alpha_{\mathrm{U-B}}<4$. Strong and long-term variability thus is generally associated with ultraviolet excess emission.

The spectral indices in the ultraviolet/blue can be directly linked to accretion via the Balmer jump flux ratio $F_U/F_B$. According to \citet{herczeg}, the threshold between accretors and non-accretors is at $F_U/F_B=0.5$, which corresponds to $\alpha_{U-B}=4.50$ (dash-dotted line in last panel of Fig. \ref{slopecorrplots}). The four objects in group A and B in our plot are on the left side of this threshold, i.e. have ongoing accretion. 

\begin{table}
\caption{Quantities derived for Fig. \ref{slopecorrplots}. \label{fig6quant}}
\center
\begin{tabular}{lcccc}
\hline
Name           & $\alpha_{3.4-4.5\,\mu m}$ &  $\alpha_{24-70\,\mu m}$ & $\alpha_{U-B}$ & $F_{1.3mm}$\\
\hline
AATau          &  -1.74   & -0.54  & 1.95    & 88.2   \\
BPTau          &  -2.76   & -1.54  &	     & 47.1   \\
CITau          &  -1.59   & -0.56  & 2.09    & 190.3  \\
CQTau          &  -0.34   &        & 3.00    & 143.3  \\
CWTau	       &  -0.21   & -1.42  & 3.58    & 96.2   \\
DFTau          &  -0.98   & -1.78  &	     &	      \\
DNTau          &  -2.68   & -0.84  &	     & 84.2   \\
DoAr24E        &   1.08   & -1.34  &	     & 51.5   \\
DOTau          &  -0.90   & -1.22  & -0.65   & 136.3  \\
DRTau          &  -0.31   &        & 3.59    & 159.3  \\
FSTau          &  -1.47   &        &	     &	      \\
FTTau          &  -1.80   & -0.60  &	     & 130.3  \\
GGTau          &  -2.72   &        &	     & 594.2  \\
GOTau          &  -2.64   & -0.22  &	     & 83.2   \\
GQLup          &  -2.09   &        &	     & 46.7   \\
Haro1-16       &  -2.46   & -0.61  &	     &  77.3  \\
Haro6-13       &  -1.16   & -1.09  &	     & 124.2  \\
Hen3-600A      &  -3.41   &        &	     &	      \\
HKTauB         &  -2.57   & -0.36  &	     &  41.1  \\
HLTau          &   3.41   &        & 3.38    & 881.7  \\
HTLup          &  -1.38   &        & 6.22    & 165.8  \\
HVTauC         &  -3.61   & -0.01  &	     &  40.1  \\
IMLup          &  -2.35   &        & 7.73    & 319.3  \\
IQTau          &  -2.28   & -0.93  &	     &  87.2  \\
IRAS04189+2650 &   3.28   &        &	     &	      \\
LkHa326        &  -1.48   & -0.89  &	     &	      \\
LkHa327        &  -0.82   & -1.46  &	     &	      \\
RULup          &  -0.40   &        &	     & 242.0  \\
RWAur          &  -0.42   &        &	     & 42.1   \\
RYTau          &   0.01   &        & 6.70    & 229.5  \\
TTauN          &   3.66   &        & 8.54    &	      \\
TWA07 	       &  -4.03   &        &	     &	      \\
TWHya 	       &  -3.84   &        &	     &	      \\
UScoJ1604-2130 &  -2.55   &        &	     &	      \\
UZTauE         &  -1.92   &        &	     & 172.3  \\
V1121Oph       &  -1.96   &        &	     & 220.8  \\
V1149Sco       &  -1.60   & -0.86  &	     &	      \\
V853Oph        &  -2.07   & -0.67  &	     &  44.2  \\
WaOph6         &  -1.84   &        &	     &  95.7  \\
\hline                          
\end{tabular}                   
\label{literature}                                             
\end{table}

\section{Correlation with infrared variability}

In Sect. \ref{iremission} it was found that the slope and maximum pooled sigma both correlate with the slope of the infrared SED between 3.4 and 4.6$\,\mu m$. Here we examine if there is also a link between these parameters and the variability at these wavelengths.

For most of our stars the WISE archive provides multi-epoch photometry at 3.4 and 4.6\,$\mu m$ (channels W1 and W2). We downloaded the WISE lightcurves from the AllWISE Multiepoch Photometry Tables. 29 stars in our primary sample are found to have useful WISE lightcurves. Typically the number of epochs per object is 20, split over two days which are separated by about 200\,d, all in 2010. A notable exception is TWHya, which has twice as many datapoints, still covering a total time window of about half a year.\footnote{We note that many of the stars in our sample have additional photometry from the post-cryo phase of WISE, for example in the NEOWISE catalogue. Due to known systematic offsets between cryo and post-cryo WISE photometry bright sources, the cryo and post-cryo cannot easily be merged. Therefore we work only with the ALLWISE datapoints.} 

As a robust metric to evaluate the variability, we calculate the reduced $\chi^2$ for all 29 stars in the two bands W1 and W2, see Table \ref{wiseresults}. All stars exhibit a reduced $\chi^2>2$ in at least one of the two bands, which is evidence for significant variability.

\begin{figure*}
\includegraphics[scale=0.45]{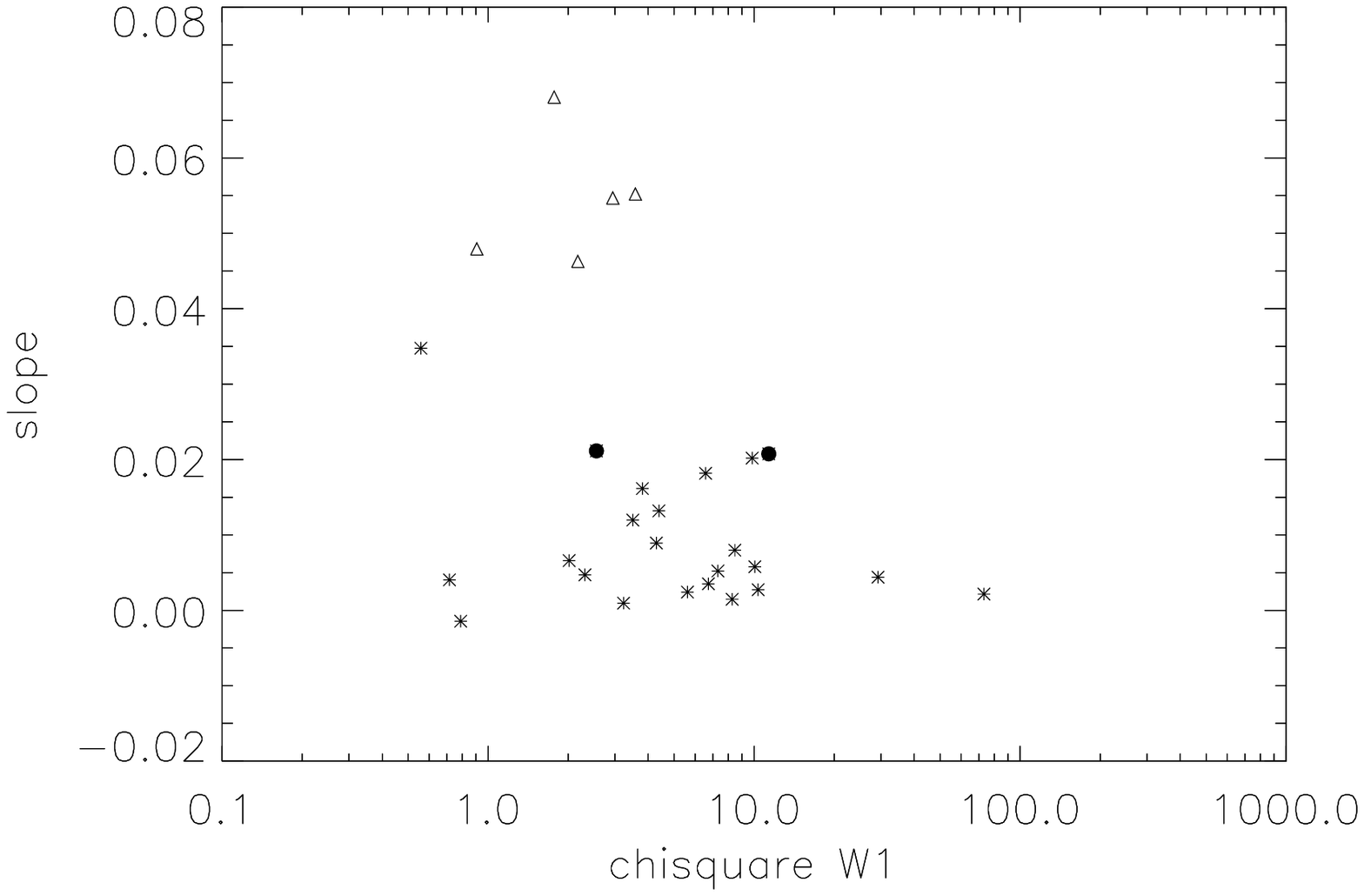}
\includegraphics[scale=0.45]{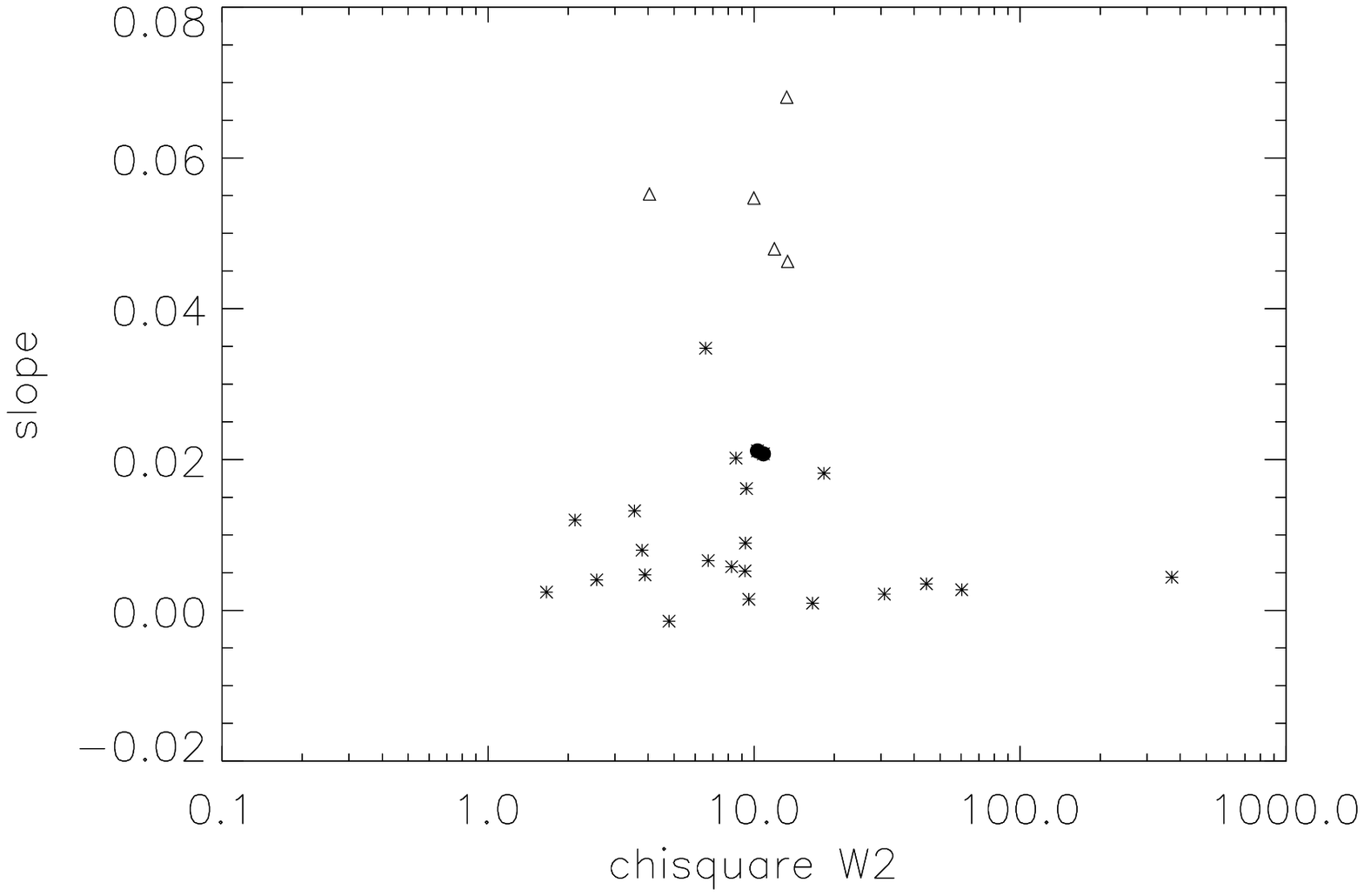}
\caption{Slope of the pooled sigma $S$ vs. infrared variability quantitied as reduced $\chi^2$, for the WISE bands W1 and W2. Stars in group A are marked in triangles, stars in group B with filled circles.}
\label{WISEvariations}
\end{figure*}

We tested for possible correlations between the infrared variability from WISE and the long-term optical variability from SWASP. The results are shown in Fig. \ref{WISEvariations}. There is no evidence for a link between these quantities. This might be explained by the fact that the two lightcurves in optical and infrared are only partially overlapping in time and are not synchronous. We also note that the WISE lightcurves do not cover the same long timescales that we cover with SWASP. Even in simultaneous optical/IR lightcurves as in the combined COROT/Spitzer campaign in NGC2264, the correlation between optical and infrared lightcurves is 'positive but weak', with frequent changes in lightcurve morphology \citep{cody14}.

\begin{figure}
\includegraphics[scale=0.45]{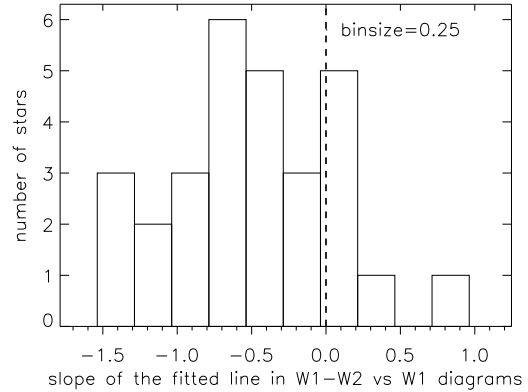}
\caption{Slope of the (W1-W2, W1) colour-magnitude trend for all stars with WISE lightcurves.}
\label{WISEslope}
\end{figure}

The multi-band photometry from WISE also gives us an opportunity to examine the colour variability in the infrared. For each star, we fit the slope in the (W1-W2,W1) colour-magnitude diagram with a straight line. The results are plotted in Fig. \ref{WISEslope} as a histogram and listed in Table \ref{wiseresults}. All stars with negative slopes in this plots become redder as they get fainter, i.e. the variability is larger in W1. The color variability in the infrared again does not show any significant correlations with the parameters of the optical variability. The observed colour variability is inconsistent with variable extinction -- with a standard extinction law \citep{yuan} the slope would become $-4.75$, whereas the largest negative value in our sample is $-1.5$. Plausible explanations for the mid-infrared variations are changes in the dust temperature, inner disk radius, or scaleheight at the inner edge of the disk. These variations could be a function of azimuth, inducing variability only through the rotation of the disk, or actual time-variability of the dust properties in the inner disk.

\begin{table}
\caption{Results of the analysis of the WISE infrared variability: Number of datapoints $N$ and $\chi^2$
in channels W1 and W2, slope $q$ in (W1-W2,W1) colour-magnitude diagram. \label{wiseresults}}
\center
\begin{tabular}{lrrrrr}
\hline
Name           & $N$ & $\chi^2_{W1}$ & $N_{W2}$ & $\chi^2_{W2}$ & $q$ \\
\hline
AATau	      &  21 &  3.58   &  21  & 4.04  &   -0.85 \\
BPTau	      &  20 &  4.39   &  16  & 3.55  &    -1.11 \\ 
CITau	      &  24 &  6.57   &  24  & 18.31 &   -0.51 \\ 
CQTau	      &  22 &  0.91   &  22  & 11.91 &   0.072 \\ 
CWTau	      &  27 &  2.55   &  27  & 10.30 &    0.15 \\ 
DFTau	      &  18 &  3.80   &  11  & 9.36  &    0.31 \\ 
DNTau	      &  21 &  10.06  &  21  & 8.23  &    -1.47 \\ 
DoAr24E       &  22 &  0.71   &  22  & 2.56  &   -0.35 \\ 
DOTau	      &  21 &  2.94   &  21  & 9.98  &   -0.55 \\ 
DRTau	      &  25 &  9.83   &  25  & 8.55  &   -0.78 \\ 
FSTau	      &  18 &  8.46   &  13  & 3.79  &   -0.94 \\ 
FTTau	      &  16 &  7.31   &  11  & 9.24  &   0.077 \\ 
GOTau	      &  22 &  5.62   &  22  & 1.66  &    -1.34 \\ 
GQLup	      &  24 &  3.23   &  24  & 16.58 &   -0.63 \\ 
Haro1-16      &  22 &  10.35  &  22  & 60.31 &   0.08 \\ 
Haro6-13      &  25 &  4.29   &  24  & 9.28  &   -0.48 \\ 
HKTauB	      &  20 &  2.31   &  20  & 3.89  &   -0.44 \\ 
HTLup	      &  24 &  0.79   &  24  & 4.79  &   -0.12 \\ 
IQTau	      &  23 &  11.35  &  23  & 10.84 &    -1.08 \\ 
IRAS04385+2550&  24 &  1.96   &  24  & 2.35  &   -0.67 \\ 
RULup	      &  25 &  1.77   &  23  & 13.26 &    0.14  \\
RWAur	      &  22 &  2.17   &  22  & 13.36 &   -0.39 \\ 
RYTau	      &  15 &  0.56   &   6  & 6.57  &   -0.23 \\ 
TWHya	      &  45 &  73.09  &  45  & 30.86 &    -1.54  \\
UScoJ1604-2130&  25 &  3.50   &  25  & 2.12  &   -0.76 \\ 
UZTauE	      &  26 &  29.23  &  26  & 371.75&    0.87 \\ 
V1121Oph      &  22 &  6.73   &  22  & 44.43 &   -0.22 \\ 
V853Oph       &  23 &  8.27   &  23  & 9.55  &   -0.89 \\ 
WaOph6	      &  24 &  2.02   &  24  & 6.72  &   -0.54 \\ 
\hline                          
\end{tabular}                   
\label{wisevar}                                             
\end{table}                

\section{Discussion and summary}

Most studies of the variability in T Tauri stars have focused on timescales of days to a few months, but changes on timescales significantly longer are well documented. In this paper we have systematically analysed the long-term variability for a sample of 39 T Tauri stars over timescales from 1 week to 7 years, using white light photometry from the Super-WASP exoplanet survey. We quantified the lightcurve variations in two parameters, the maximum pooled sigma over any timescale $\sigma _{max}$ and the increase of pooled sigma as a function of time $S$. The first gives an indication of the amplitude of the variability independent of timescale, the second quantifies the increase of variability over time, which is our primary interest here. We searched for periods, for correlations between variability characteristics, stellar/disk/accretion diagnostics, and infrared variability. Our study should be seen as reference study for a well-characterised sample of objects and as a step towards the full characterisation of the long-term variations in young stars. In the following we summarise and discuss our main findings.

\begin{enumerate}
\item{Most stars show evidence for variability $\sigma _{max} <0.3$\,mag, which reaches its maximum after 1-4 weeks, without evidence for further increase on longer timescales. This is consistent with variability on rotational timescales, most plausibly caused by cool or hot spots or variable extinction due to structures at the inner edge of the disk. This was already found by \citet{grankin} and is also reported in the accretion rate monitoring program by \citet{costigan}. Thus, for {\it typical} young stars a few weeks of photometric monitoring is sufficient to measure the total amount of the variations (up to timescales of 7 years) and to identify stars with anomalous behaviour. This result is relevant for the transient detection in current and future large-scale variability surveys like Gaia or LSST.}
\item{About one fifth of our sample (8/39) shows a maximum pooled sigma exceeding 0.3\,mag. Out of these eight, five also show a significant increase in the pooled sigma, which is evidence for long-term changes over timescales of months or years. As far as accretion information is available, all stars with evidence for long-term variability are accretors. The variability amplitude and its increase over time are positively correlated with the slope of the mid-infrared slope at 3-5$\,\mu m$. Thus, the long-term optical variability in T Tauri stars is related to accretion and the properties in the inner disk.}
\item{Four out of the five objects with long-term variability show evidence for temporary cyclic behaviour with periods of 20-60 days, significantly exceeding the rotation period. One option for these long-term cycles is periodic obscuration by features in the inner disk. Assuming a disk in Keplerian rotation, a periodicity of 20-60\,d translates into a distance from the star of 0.1-0.3\,AU, for typical stellar masses in our sample. For comparison, the dust sublimation radius for parameters typical in our sample is located between 0.03 and 0.04\,AU away from the star.\footnote{This can be calculated using equation (1) in \cite{monnier}, adopting $Q_R\sim1$ and $T_S\sim1500K$.} Thus, it is conceivable that the periods arise from dusty structures in the inner parts of the disk. However, obscurations are only expected for disks seen at high inclination, and two of the objects with long periods have $i<70$\,deg. A second option to explain the long-term periods are instabilities in the magnetospheric accretion from the inner disk onto the star. It is expected that magnetic field lines are twisted as a result of differential rotation. This may lead to opening and later reconnection of magnetic flux tubes, modulating the accretion flow onto the central object and thus the excess emission due to accretion. This behaviour might be expected to repeat itself with a typical cycle of a few rotation periods, possibly giving rise to brightness variations with periods as observed in our sample. This scenario has already been discussed by \cite{bouvier2003} to explain the detailed observational picture of AATau. Our results might be evidence for such behaviour in a larger group of objects.}
\end{enumerate}

\section*{Acknowledgments}
The research leading to these results has received funding from the European Union Seventh Framework Programme
FP7-2011 under grant agreement no 284405.
WASP-South is hosted by the South African Astronomical Observatory and SuperWASP-North by the Isaac Newton Group and 
the Instituto de Astrofisica de Canarias; we are grateful for their ongoing support and assistance. Funding for 
WASP comes from consortium universities and from the UK's Science and Technology Facilities Council.
This publication makes use of VOSA, developed under the Spanish Virtual Observatory project supported from the 
Spanish MICINN through grant AyA2008-02156. 

\newcommand\aj{AJ} 
\newcommand\actaa{AcA} 
\newcommand\araa{ARA\&A} 
\newcommand\apj{ApJ} 
\newcommand\apjl{ApJ} 
\newcommand\apjs{ApJS} 
\newcommand\aap{A\&A} 
\newcommand\aapr{A\&A~Rev.} 
\newcommand\aaps{A\&AS} 
\newcommand\mnras{MNRAS} 
\newcommand\pasa{PASA} 
\newcommand\pasp{PASP} 
\newcommand\pasj{PASJ} 
\newcommand\solphys{Sol.~Phys.} 
\newcommand\nat{Nature} 
\newcommand\bain{Bulletin of the Astronomical Institutes of the Netherlands}
\newcommand\memsai{Mem. Soc. Astron. Ital.}
\newcommand\apss{Astrophysics and Space Science}

\bibliographystyle{mn2e} 
\bibliography{biblio}
\end{document}